\documentclass[twocolumn]{article}

\usepackage[utf8]{inputenc}
\usepackage[T1]{fontenc}
\usepackage{booktabs}
\usepackage[hmargin=2cm]{geometry}

\usepackage{graphicx}
\usepackage{balance}  % for  \balance command ON LAST PAGE  (only there!)

\usepackage[utf8]{inputenc}
\usepackage[T1]{fontenc}
\usepackage{mathptmx}
\usepackage{caption}
\usepackage{subcaption}
\usepackage{float}
\usepackage[vlined,figure]{algorithm2e}
\usepackage{amssymb}
\usepackage[table]{xcolor}
\usepackage{authblk}
\usepackage{tikz}
\usepackage{pgfplots}
\usepackage{hyperref}

\SetFuncSty{textsc}
\newcommand{\api}[1]{\textsc{#1}}
\usepgfplotslibrary{groupplots}
\usetikzlibrary{patterns}
\definecolor{darkcandyapplered}{rgb}{0.64, 0.0, 0.0}
\definecolor{darkcoral}{rgb}{0.8, 0.36, 0.27}
\definecolor{darkcyan}{rgb}{0.0, 0.55, 0.55}
\definecolor{darkgoldenrod}{rgb}{0.72, 0.53, 0.04}
\title{Lazy State Determination:\\More concurrency for contending linearizable transactions} 

\author[1]{Tiago M. Vale}
\author[1]{João Leitão}
\author[1]{Nuno Preguiça}
\author[2]{Rodrigo Rodrigues}
\author[1]{Ricardo J. Dias}
\author[1]{João M. Lourenço}
\affil[1]{%
       NOVA LINCS, DI, FCT\\
       Univerisdade NOVA de Lisboa
}
\affil[2]{%
       INESC-ID, IST\\
       Universidade de Lisboa
}

\date{June, 2020}

\begin{document}
\maketitle

\begin{abstract}
The concurrency control algorithms in transactional systems limits concurrency to provide strong semantics, which leads to poor performance under high contention.
As a consequence, many transactional systems eschew strong semantics to achieve acceptable performance.
We show that by leveraging semantic information associated with the transactional programs to increase concurrency, it is possible to significantly improve performance while maintaining linearizability.
To this end, we introduce the lazy~state~determination API to easily expose the semantics of application transactions to the database, and propose new optimistic and pessimistic concurrency control algorithms that leverage this information to safely increase concurrency in the presence of contention.
Our evaluation shows that our approach can achieve up to $5\times$ more throughput with $1.5\times$ less latency than standard techniques in the popular TPC-C benchmark.
% , with only minor changes to the existing codebase.
\end{abstract}

\section{Introduction} % (fold)
\label{sec:introduction}
Linearizable transactions provide a simple and powerful abstraction to programmers: transactions appear to complete atomically, one at a time despite executing concurrently.
This property greatly simplifies developing and reasoning about concurrent applications.

In recent years, we saw a continuing interest in research on transactional systems, e.g., as transactional properties were adopted in ``NoSQL'' systems~\cite{dynamo-sosp-2007}, or as the performance of distributed transactions was improved through new hardware features~\cite{farm-sosp-2015}.
However, this research does not fundamentally improve the performance of transactions in the presence of their Achilles heel: contention~\cite{stonebraker:blog}. When transactions conflict with one another, they end up executing most of their logic one at a time. To circumvent this limitation, transactional systems may resort to weaker semantics, but these often break the integrity of the applications that are built using such transactions~\cite{bailis:feral}.

In this paper, we show that it is possible to achieve linearizable transactions with significantly better performance than existing techniques by building on the observation that the lack of semantic information about the transaction leads to a conservative view of what is a conflict, and therefore imposes unnecessary synchronization between transactions.
For example, two transactions that increase the number of items in an inventory will be treated as conflicting because they both write to the database tuple containing the total quantity.
However, the semantics of those transactions do not imply a conflict, provided that the aggregated effects of both transactions are applied to the database.

To address this shortcoming, we propose lazy~state~determination~(\api{lsd}), a novel API for defining transactions that conveys their semantics to the database. 
The main insight behind \api{lsd} is that by exploring the semantics of the transaction, it is possible to increase concurrency while still providing linearizability~\cite{lineariz} (also known as strict~serializability~\cite{serializability-1979}). 
This contrasts with previous work that explores semantic information to improve transaction processing~\cite{escrow-tods-1986,homeostasis-sigmod-2015,indigo-eurosys-2015}, which focuses on maintaining specific application invariants under consistency models weaker than serializability. 

One important challenge in our work is how to expose the semantic information without requiring programmers to significantly modify their coding practices, or make significant changes to existing applications.
To this end, we realize \api{lsd} by having the \api{read} operation return a future~\cite{futures-1977} (an opaque proxy for a value) instead of a concrete value, and materializing futures as late as possible, i.e., only when the transaction commits.
To allow transactions to still be expressive with futures without resolving them, we:~(a)~introduce a new operation, \api{is-true}, that allows transactions to specify conditions over futures, and~(b)~provide operations that allow transactions to specify their updates to the database as lazily-evaluated functions that can use futures.
This differs from prior works~\cite{warranties-nsdi-2014,sloth-sigmod-2014} that need to materialize \emph{delayed} reads as soon as they are used by the transaction logic. 
%Warranties~\cite{warranties-nsdi-2014} also allow transactions to express conditions that must hold for the transactions to succeed, but cannot increase concurrency for transactions that perform computations using the values read or that externalize them.
%Additionally, unlike systems~\cite{callas-sosp-2015} that require the full set of transactions to be known \emph{a priori}, it poses no requirement for adding new functionality to the system. 

This novel API allows \api{lsd} transactions to execute over an abstract database state, and resolve this abstract state as late as possible, thus increasing the chances for safely committing without breaking isolation. 
To demonstrate this, we modified existing optimistic and pessimistic concurrency control protocols to allow for conditional~validation.
The key idea of this design is to verify that the required conditions still hold when the transaction attempts to commit (in the case of optimistic concurrency control), or to use a condition~lock acquired in condition~mode for a certain condition~$c$, which is only compatible with an acquisition in write~mode if the value that will be written respects the condition~$c$ (in the case of pessimistic concurrency control).

We implemented and evaluated a prototype transactional key-value store that provides linearizable transactions using the \api{lsd} interface.
Our evaluation shows that \api{lsd} transactions achieve up to~$5\times$ more throughput with~$1.5\times$ less latency than standard transactions under high contention in our experiments with the popular TPC-C benchmark~\cite{tpcc}.

In summary, we make the following contributions:

\paragraph{LSD API.}
We propose \api{lsd}, an interface to express transactions that allows the database to collect semantic information useful to achieve higher performance under contention. 

\paragraph{Concurrency control.}
We propose new optimistic and pessimistic concurrency control algorithms for providing linearizability while exploring semantic information to increase concurrency in the presence of contention, using novel \emph{condition~validation} and \emph{condition~locking} techniques.

% \noindent{\textbf{\api{lsd} and 2PC.}}
% We describe how to adapt 2-phase~commit~(2PC~\cite{databasebook-1987}) to take advantage of \api{lsd}, and discuss \api{lsd}'s impact on distributed transactions.

\paragraph{Evaluation.}
We implemented an \api{lsd}-compatible prototype and evaluated it using TPC-C and microbenchmarks, showing significant performance improvements.

The remainder of this paper is organized as follows.
We present an overview of \api{lsd} in~\S\ref{sec:overview}, by motivating the shortcomings of the standard interface through an example, from which we derive the \api{lsd} interface.
\S\ref{sec:design} then presents \api{lsd} in detail, including the design of \api{lsd}-aware variants of OCC and 2PL.
% in~\S\ref{sub:design_cc}.
% , and how to adapt 2PC for distributed \api{lsd} transactions in~\S\ref{sub:design_dtxns}.
\S\ref{sec:evaluation} describes our prototype and the results of our evaluation.
\S\ref{sec:related_work} discusses \api{lsd} in the context of related work and \S\ref{sec:conclusion} concludes the paper.
% section (end)
\section{Overview} % (fold)
\label{sec:overview}
A typical database API exposes five operations: (1)~\api{begin}: starts a new transaction, (2)~\api{read}($key$): returns the value of the database object identified by~$key$, (3)~\api{write}($key$, $val$): modifies the value of the object identified by~$key$ to~$val$, (4)~\api{commit}: commit the current transaction, and (5)~\api{abort}: aborts the current transaction.

Conceptually, a transaction is a function~$f$ that changes the data\-base from an initial state~$s_i$ to a final state~$s_f$, i.e. $f(s_i) = s_f$.
In light of this formulation, the \api{read} and \api{write} API calls allow transactions to specify the final state~(\api{write}) as a function of the initial state~(\api{read}).

\begin{figure}
    \begin{subfigure}[b]{0.47\linewidth}
        \begin{algorithm}[H]
\small
\DontPrintSemicolon
\SetInd{0.1em}{0.5em}
\SetKwFunction{TMbegin}{begin}
\SetKwFunction{TMread}{read}
\SetKwFunction{TMis}{is-true}
\SetKwFunction{TMwrite}{write}
\SetKwFunction{TMcommit}{commit}
\SetKwFunction{TMabort}{abort}
\SetKwProg{Function}{upon}{}{}

\TMbegin\;
$v$ $\gets$ \TMread{$stock$}\;
\eIf{$v \ge qty$}{
    $v$ $\gets$ $v - qty$\;
    \TMwrite{$stock$, $v$}\;
    \TMcommit\;
}{
    \TMabort\;
}
        \end{algorithm}
        \caption{Traditional interface.}
        \label{subfig:api_traditional}
    \end{subfigure}
    \begin{subfigure}[b]{0.51\linewidth}
        \begin{algorithm}[H]
\small
\DontPrintSemicolon
\SetInd{0.1em}{0.5em}
\SetKwFunction{TMbegin}{begin}
\SetKwFunction{TMread}{read}
\SetKwFunction{TMis}{is-true}
\SetKwFunction{TMwrite}{write}
\SetKwFunction{TMcommit}{commit}
\SetKwFunction{TMabort}{abort}
\SetKwProg{Function}{upon}{}{}

\TMbegin\;
$\Box$ $\gets$ \TMread{$stock$}\;
\eIf{\TMis{$\{\Box \ge qty\}$}}{
    $\bigtriangleup$ $\gets$ $\{\Box - qty\}$\;
    \TMwrite{$stock$, $\bigtriangleup$}\;
    \TMcommit\;
}{
    \TMabort\;
}
        \end{algorithm}
        \caption{\api{lsd} interface.}
        \label{subfig:api_lsd}
    \end{subfigure}
    \caption{Simplified portion of TPC-C's New~Order-like transaction.}
    \label{fig:api}
\end{figure}

% \subsection{The pitfalls of the traditional API} % (fold)
% \label{sub:overview_traditional_api}
\paragraph{The pitfalls of the traditional API.}% (fold)
Consider the example in Figure~\ref{subfig:api_traditional} that depicts a simplified portion of the TPC-C new order transaction~\cite{tpcc}, which implements the action of buying a certain quantity~$qty$ of items.
If the item's stock~$\big($$v \gets$ \api{read}($stock$)$\big)$ is enough to fulfil the order~$\big($$v \ge qty$$\big)$, the stock value decreases by~$qty$~$\big($\api{write}($stock$, $v - qty$)$\big)$.

This example illustrates how the~\api{read}/\api{write} interface fails to convey the semantics of the transaction to the database, e.g., the dependencies of the transaction behavior on the values it reads, or how it computes the values it writes.
From the point of view of the database, transactions are a sequence of opaque~\api{read}/\api{write} operations.
(This is true regardless of whether transactions execute co-located with the database, as stored procedures, or in a remote client.)

To understand how this can be a limiting factor, consider the situation where the current stock value is~42~($stock$), and the quantity to order is~1~($qty$).
When the transaction issues the \api{read}({$stock$}) operation, the database returns the value~42.
Since the database does not know what the transaction will do with the returned value, it must be conservative to account for all possible situations, e.g., the transaction only executing some operations depending on the returned value, or using the returned value to perform a computation that returns the value of a subsequent~\api{write}.
As a consequence, 2PL must lock the~$stock$ object to prevent any other transaction from modifying it and invalidate any branching decision or computed value by the transaction that observed the value~42.
Similarly, OCC records the read operation so that the database can check that the~$stock$'s value is the same when the transaction attempts to commit; if meanwhile another transaction modifies the~$stock$, transactions that observed the now-stale stock value fail to commit.

As this example shows, a central part of enforcing transaction isolation is ensuring that the state that a transaction observes (i.e., the values returned by \api{read} operations) remains unchanged throughout its execution.
Our key insight is to question whether a transaction \emph{really} needs to observe a specific state during its execution.
In other words, in our running example, does the~\api{read}($stock$) operation really need to expose a particular state to the transaction before commit (e.g., the value 42)?
With the current interface the answer is yes.
Otherwise, transactions cannot have conditional branches that depend on the database state, nor perform updates to the state that are a function of that state.
Going back to our example, the transaction could not check whether there is enough stock nor compute the new stock value.
% subsection (end)

% \subsection{Introducing LSD} % (fold)
% \label{sub:overview_lsd}
\paragraph{Introducing LSD.} % (fold
In this paper, we overcome these limitations by rethinking the transactional API in order to provide linearizable transactions that allow for greater concurrency.
The key observation behind \api{lsd} is that, in general, transactions do \emph{not} need to observe a concrete state to execute most of their logic.
Thus, we propose alternative semantics for the~\api{read} operation.
Specifically, the~\api{read} operation should \emph{not} expose a specific database state by returning a concrete value, but should instead return a \emph{future}~\cite{futures-1977}.

A future is an object that acts as a proxy for a value that is initially unknown.
In our case, a future symbolizes the value of a specific database object.
This means that the database \emph{promises} to resolve the future's value, but does not do it right away.
In particular, we want to defer evaluating futures until the transaction attempts to commit~(\emph{lazy evaluation}~\cite{lazyeval-acmcsur-1989}) to maximize concurrency.
(Note that the traditional semantics of the~\api{read} operation is equivalent to returning futures that are immediately resolved.)
Returning to our running example, we depict this modification in Figure~\ref{subfig:api_lsd} with the future that symbolizes the stock value as~$\Box$.

The proposed change to the semantics of the~\api{read} operation has a clear benefit: if a transaction does not observe a specific state, other transactions can modify it without breaking the isolation guarantees of the first transaction.
However, this raises the problem of determining how can a transaction use futures.
This can, in turn, be split into two main challenges.
The first is how can a transaction perform conditional branching based on futures.
The second is how can a transaction compute values that depend on futures.
For instance, how can the logic of our example transaction decide whether it can fulfil the order if it does not know the stock value, and how can the transaction compute the new stock value?
(A naive approach is to eagerly resolve futures when a transaction requires their value, but this restricts concurrency.)

To solve the first challenge, we observe that a future symbolizes the value of a particular database object.
While we would like that a transaction is not able to directly observe the value of a future, we can still ask the database whether a future's value respects a certain condition.
For example, the transaction can ask the database whether the stock value is greater than~$qty$, and make a control flow decision depending on the database's answer.

To support this functionality we introduce a new operation,~\api{is-true}($c$), which, given a condition~$c$ over one (or more) futures, returns whether the condition holds or not.
We show the~\api{is-true} operation using the~${\Box \ge qty}$ condition in Figure~\ref{subfig:api_lsd}.
Note that while the~\api{is-true} operation effectively exposes database state to the transaction, it exposes an abstract state~(the stock is greater than~$qty$) rather than a concrete one~(the stock is~$42$), which has the potential to allow for more concurrency, e.g., by allowing concurrent modifications of the stock value as long as it retains a non-negative value after all the modifications.

The second challenge is how can a transaction perform computations using futures.
To solve this challenge, we observe that while a transaction cannot perform the actual computation with futures, it can define the necessary computation and let the database perform it when the transaction commits and the futures are resolved to concrete values.
For example, the transaction can define that the new stock value is whatever value its future ends up resolving to minus~$qty$.

To support this behavior we change the semantics of the~\api{write} operation so that, instead of receiving the concrete new value for an object, it receives a function that computes the concrete value when evaluated.
This function has the important property that it can depend on the values of any future, since the database can resolve them.
Furthermore,~\api{write} functions are lazily evaluated by the database when the transaction commits, so that the futures that the functions depend on may remain unresolved.
In Figure~\ref{subfig:api_lsd}, we represent this function as~${\{\Box - qty \}}$, which is the argument of the~\api{write} operation.

We expect that the proposed changes to the \api{read} and \api{write} operations and the addition of the~\api{is-true} operation will enable the database to provide linearizable transactions with more concurrency, potentially resulting in higher throughput and lower latency.

On the one hand we decrease the time window in which a transaction requires isolation.
With the traditional interface, the transaction requires isolation from the moment when it first observes database state (with the traditional \api{read} operation) until the transaction attempts to commit.
With \api{lsd} the transaction only requires isolation during its commit operation if it does not require any specific conditions.

On the other is that we reduce the set of concurrent transactions that are forced to abort/wait when executing concurrently with some transaction to guarantee the required isolation level.
Even when a transaction needs to test some condition over database objects, \api{lsd}'s~\api{is-true} operation still allows concurrent transactions to modify those objects as long as these modifications do not invalidate the previously asserted conditions.
This contrasts with the traditional interface that prevents any modifications, whether they violate such conditions or not.
This leads to lower abort rates/waiting, and hence to a higher amount of useful work.

That said, the \api{lsd} API is not a panacea.
Transactions that must observe a concrete state can not reap \api{lsd}'s benefits.
For example, transactions that externalize values during their execution need to resolve the required futures, falling back to the standard \api{read} semantics.
However, we believe that a large class of transactions can take advantage of \api{lsd} proposed semantics.
% subsection (end)
% section (end)
\section{LSD Design} % (fold)
\label{sec:design}
% \subsection{Goal} % (fold)
% \label{sub:design_goals}
The high-level goal of \api{lsd} is to allow databases to provide linearizable transactions with higher performance than what can typically be achieved, while minimizing changes in the way programmers specify the logic of their transactions.
%To achieve this goal, we revisit the interface that the database exports to allow applications to execute transactions against it, which, as mentioned, traditionally consists of the \api{begin}, \api{read}, \api{write}, \api{commit}, and \api{abort} operations.
%The key observation that drives our design is that the performance of ACID transactions is affected by the need to ensure isolation between concurrent transactions, and yet isolation is often only required when transactions observe a concrete database state.
%Therefore, we propose semantic changes to the \api{read} and \api{write} operations, and a new \api{is-true} operation, to allow transactions to execute their logic, to the extent possible, on an abstract database state (preferably until the point when they attempt to \api{commit}).
%This way, we delay the overhead of maintaining isolation and centralize it in the \api{commit} operation, thus reducing the interval during which the database needs to provide isolation, and consequently improving performance.
% subsection (end)
\subsection{Design overview} % (fold)
\label{sub:design_overview}
\begin{figure}[tb]
    \centering
	\small
    \includegraphics[width=0.8\linewidth]{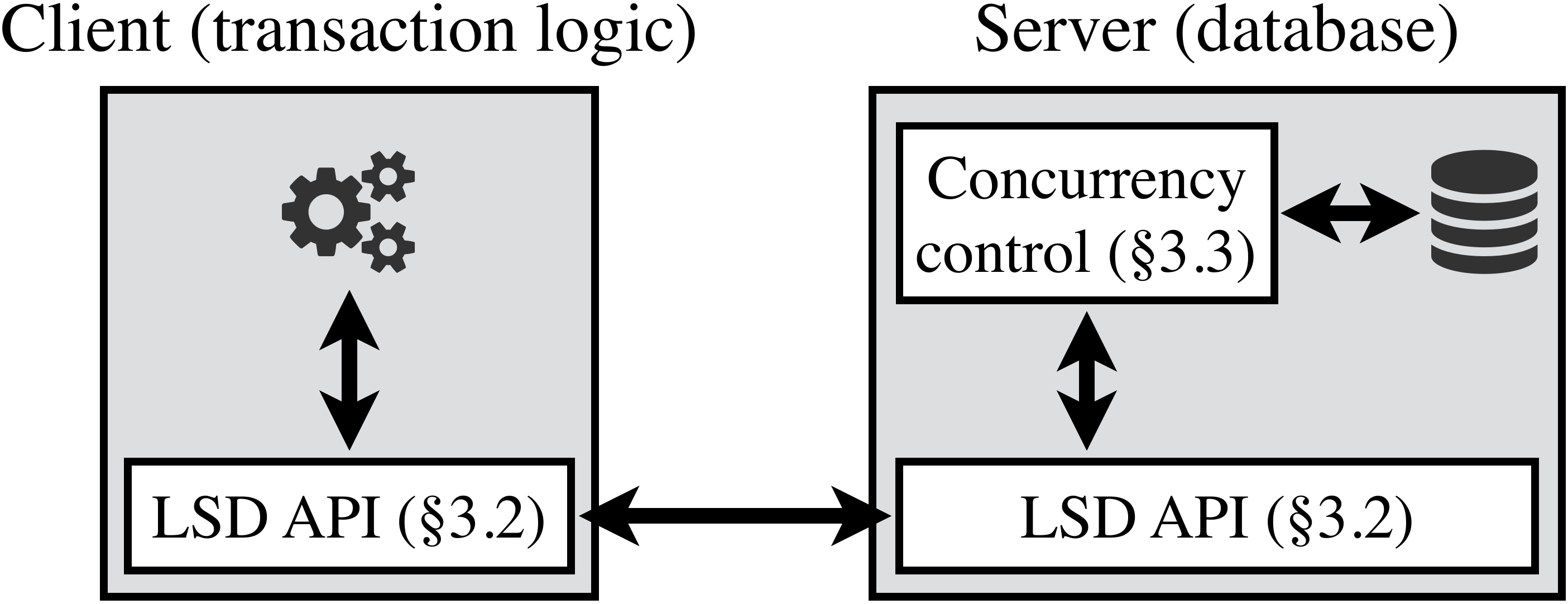}
    \caption{Overview of the system's architecture.}
    \label{fig:design_overview}
\end{figure}
Figure~\ref{fig:design_overview} shows the main components of our design.
Clients execute application code that interacts with the database server via transactions written using the \api{lsd} API.
Note that these are logical components, meaning that our design does not make assumptions regarding the physical relationship between clients and servers, nor the physical realization of the server.
For example, clients can be physically separated from the server or co-located with it (e.g., in a stored procedure), and the database may or may not be partitioned or replicated.
Nevertheless, for the rest of this paper we assume that clients execute transactions and are separated from the server, which is the case in our prototype and evaluation.
% subsection (end)
\subsection{Interface} % (fold)
\label{sub:design_interface}
\begin{figure*}[tb]
    \centering
    \small
    \begin{tabular}{l|p{11.75cm}}
\api{Operation} & \api{Description} \\
\midrule
\api{begin} & Starts a new transaction. \\
\api{read(}\emph{key}\api{)} $\rightarrow$ $\Box$ & Returns $\Box$, a future for the value of object \emph{key}. \\
\api{read(}$\bigtriangleup$\api{)} $\rightarrow$ $\Box$ & Evaluates the future~$\bigtriangleup$ and returns~$\Box$, a future for the value of the object that~$\bigtriangleup$ evaluates to. \\
\api{is-true(}$\Box$\api{)} $\rightarrow$ boolean & Returns whether the condition~$\Box$ is currently true in the database. \\
\api{write(}\emph{key}\api{,} $\Box$\api{)} & Updates object \emph{key}'s value to the value that~$\Box$ will evaluate to. \\
\api{write(}$\bigtriangleup$\api{,} $\Box$\api{)} & Updates the value of the object that~$\bigtriangleup$ will evaluate to, to the value that~$\Box$ will evaluate to. \\
\api{commit} $\rightarrow$ boolean & Attempts to commit the ongoing transaction. \\
\api{abort} & Aborts the ongoing transaction. \\
    \end{tabular}
    \caption{The \api{lsd} interface operations. The symbols $\Box$ and $\bigtriangleup$ denote futures.}
    \label{fig:lsd_api}
\end{figure*}
Figure~\ref{fig:lsd_api} shows the \api{lsd} interface, which allows applications to execute transactions against the database.
The~\api{begin},~\api{commit}, and~\api{abort} operations are the standard operations.
They allow an application to start, commit, and abort a transaction, respectively.
\api{lsd} introduces two changes to the standard interface: new semantics for the~\api{read} and~\api{write} operations, and a new~\api{is-true} operation.
We first describe the new~\api{read} and~\api{write} operation semantics, then present the~\api{is-true} operation, and finally address the case when a transaction wants to access an unknown object, i.e., the object identifier is itself a future.

\paragraph{\api{read}.}
The typical semantics of the~\api{read} operation is to return the current value of the object, which requires the concurrency control protocol to kick in as a result of exposing the database state to transactions.
In contrast, \api{lsd}'s~\api{read} operation returns a future for the value of a given object, instead of exposing the object's current concrete value.
From the application's point of view, this future is an opaque representation of the object's value.
However, the database knows how to interpret such future; in particular, it has the possibility to \emph{resolve} the future, i.e., compute the value that the future represents, which is to \emph{actually} read and return the object's value.
Thus, informally the contract that \api{lsd} provides between the transaction and the database is the following: the transaction should use the future as if it is the actual value, and the database promises to lazily resolve the future such that, when the transaction commits, it is as if it executed with the concrete value instead of the future.
The benefit of these semantics is that the concurrency control protocol only needs to intervene when the database resolves a future and \emph{not} when a transaction issues a~\api{read} operation.

\paragraph{\api{write}.}
The traditional~\api{write} operation receives both the identifier of the object and its new value.
This interface fits well with the traditional~\api{read} operation since reads return concrete values, so if a transaction wants to modify the value of an object it can read the object, compute the modified value, and write this new value.
However, since the \api{lsd}~\api{read} operation returns a future instead of a concrete value, the transaction should be able to modify and write values derived from futures, instead of concrete values.
To address this, we have two choices.
The first is to resolve the future so that the transaction can perform its modification.
This approach goes against \api{lsd}'s goal, since resolving futures exposes database state to transactions, which in turn requires the concurrency control algorithm to enforce the required isolation.
The second choice, which we follow, is \emph{defining but not performing} the computation necessary to modify the value, so that futures may remain unresolved to promote parallelism.
To do so, the transaction specifies the computation it needs to do as a function that, when evaluated by the database, computes the new value for the object.
For instance in our running example of Figure~\ref{subfig:api_lsd}, where a transaction wants to decrease the avaliable stock for a given item, the transaction reads the stock and obtains $\Box$~(its future value), and defines the function that decreases the stock~($\Box - qty$).
This function is also a future: it represents the value that the transaction intends to write to the stock object.
For this approach to work, the database needs to know how to evaluate such functions so that, when the transaction commits, the database can install the object's new value.
To understand how this can be done, we observe that we can divide this function evaluation into two parts: resolving future reads on which the function depends on, and executing the function's logic.
As discussed, the database knows how to resolve future reads.
As for executing the function's logic, the idea is that we define this in a way that the database can initially refer to the function without resolving it, but at commit time interpret and execute it.
To achieve this, in our prototype, we provide transactions with a library of operations, which can be composed to create functions, e.g., \api{sub(}$\Box$\api{,} \emph{qty}\api{)} to decrease the stock in our example. 

%\comment[rodrigo]{Seria bom referenciar uma figura com a API completa em vez de dar so um exemplo.}

\paragraph{\api{is-true}.}
So far, we managed to prevent exposing data\-base state to transactions by changing the semantics of the \api{read} and \api{write} operations.
However, transactions may need to decide what to do based on the database state, as exemplified in our running example where the transaction only orders the item if there is enough stock available.
As before, we want to avoid resolving the futures required to make the decision of what to do, we introduce the~\api{is-true} operation, which, given a condition over the database state, returns whether the condition holds or not.
This condition is a function that, as discussed for the~\api{write} operation, can depend on futures.
%\comment[rodrigo]{As frases seguintes tem falta de coesao, cada frase faz sentido individualmente mas as frases nao ligam umas com as outras e o texto torna-se dificil de ler e compreender.}
In our running example, the transaction decides to decrease the stock depending on whether there is enough stock available:~${\Box \ge qty}$.
(In our prototype, we also provide transactions with operations to create conditions, e.g., \api{gte(}$\Box$\api{,} \emph{qty}\api{)}, to check whether there is enough stock.)

Note that the~\api{is-true} operation \emph{does} expose database state to transactions, but this is inevitable if the transaction performs different actions depending on the database state.
The merit of the \api{is-true} operation is that it exposes \emph{abstract}, instead of \emph{concrete}, state to transactions, which enables the database to maintain isolation while potentially allowing more parallelism.
For instance, if the transaction of our running example attempts to purchase a quantity of 4, and the current stock is 42, the~\api{is-true} operation returns $\top$.
Other concurrent transactions may successfully update the stock value and commit without breaking isolation as long as the stock value remains greater or equal to 4.
Enforcing the semantics of this operation requires the concurrency control protocol to either: (1)~ensure that the result of the~\api{is-true} operation remains valid until the transaction commits (pessimistic approach, 2PL-style), or (2)~abort the transaction when it attempts to commit if the result of the condition no longer holds (optimistic approach, OCC-style).
In the next section we discuss how to adapt both 2PL and OCC to support for the~\api{is-true} operation.
We implemented both approaches in our prototype.

\paragraph{Futures as keys.}
Up until this point, we have not discussed what happens when the transaction attempts to read or write an object whose identifier is itself a future.
For reads, this situation is likely to happen when accessing objects via a secondary index.
Secondary indexes are seldom kept on keys whose values are updated frequently since they tend to be expensive to modify~\cite{calvin-sigmod-2012}.
%Therefore, when \api{lsd}'s~\api{read} operation receives a future as a parameter, we resolve the future immediately in order to know which object is being read.
Given this observation, we chose to resolve the future immediately when a \api{read} operation receives a future as a parameter, in order to know which object is being read. This simplifies reasoning and implementation effort, since the alternative of maintaining ``futures of futures'' would require a chain of resolves at commit time.
As for future identifiers in~\api{write} operations, we chose to keep them unresolved because transactions may write to objects whose future-keys depend on the database state.
This is the case, for example, when assigning unique identifiers to keys from a monotonically increasing counter, which we believe to be a common programming idiom.\footnote{This is the case in the popular TPC-C benchmark~\cite{tpcc}. The fact that purchase orders have monotonically increasing identifiers not only guarantees uniqueness, but also serves to identify and compare the recency of each order.}
As such, if we resolve the future identifier immediately, we risk exposing highly-contended database state to transactions, which goes against our design goals.
The price we pay for our decision is that, in the general case of distributed transactions, they may require an additional communication round with servers to commit.
We discuss this aspect further in \S\ref{sub:design_dtxns}.
% subsection (end)
\subsection{Concurrency control} % (fold)
\label{sub:design_cc}
Now we turn our attention to the impact the \api{lsd} API has on concurrency control, and discuss how to adapt two popular concurrency control protocols: OCC~\cite{occ-1981,silo-sosp-2013} and 2PL~\cite{2pl-1976,databasebook-1987}.
The two main elements of the \api{lsd} API that drive the adaption are:~(1)~futures, as the protocol needs to be aware of them to know what to do at commit time, and~(2)~the~\api{is-true} operation, which exposes abstract database state to transactions and therefore requires concurrency control.

\subsubsection{Overview} % (fold)
\label{subsub:design_cc_overview}
The high level idea of the adaptation of both OCC and 2PL is to maintain two extra read and write~sets, which we call \emph{future~read~and~write~sets}, to keep futures unresolved until commit time, and a \emph{condition~set} to support the \api{is-true} operation and conditions.
The \api{lsd}-aware OCC and 2PL protocols differ mainly on how they handle the condition~set.
OCC verifies that the conditions still hold at commit time while 2PL ensures that concurrent transactions that write values that invalidate active conditions cannot commit while such conditions are active.
%Otherwise, the new protocols follow the essence of the original protocols.

Figures~\ref{fig:clt_occ_lsd} and~\ref{fig:clt_2pl_lsd} show the \api{lsd}-aware OCC and 2PL protocols, respectively.
The behavior of the \api{begin}, \api{read(}$key$\api{)}, and \api{write} operations is protocol-agnostic so we start by describing these before detailing the protocols for OCC~(\S\ref{subsub:design_cc_occ}) and 2PL~(\S\ref{subsub:design_cc_2pl}).

\paragraph{\api{begin}.}
Initializes the read/write~sets, future read/write~set, and condition~set (\emph{rset}, \emph{wset}, \emph{frset}, \emph{fwset}, and \emph{cset}, respectively.)

\paragraph{\api{read(}$key$\api{)}.}
Creates a future-value for $key$'s value~($\Box$), add it to the future~read~set, and returns it. (This is a local operation).

\paragraph{\api{write(}$key$\api{,}$\Box$\api{)}.}
Buffers~$\Box$, the future-value for~$key$, in the write~set.

\paragraph{\api{write(}$\bigtriangleup$\api{,}$\Box$\api{)}.}
Buffers~$\Box$, the future-value to assign the future-key~$\bigtriangleup$, in the future~write~set.
% subsubsection (end)
\subsubsection{Optimistic concurrency control (OCC)} % (fold)
\label{subsub:design_cc_occ}

\begin{algorithm}[htp]
\linespread{1.15}\selectfont
\small
\DontPrintSemicolon
\SetInd{0.1em}{0.5em}
\SetKwFunction{TMbegin}{begin}
\SetKwFunction{TMread}{read}
\SetKwFunction{TMis}{is-true}
\SetKwFunction{TMwrite}{write}
\SetKwFunction{TMcommit}{commit}
\SetKwFunction{TMabort}{abort}
\SetKwFunction{Future}{future}
\SetKwFunction{Keys}{keys}
\SetKwFunction{Key}{key}
\SetKwFunction{Resolve}{resolve}
\SetKwFunction{DMread}{dm-read}
\SetKwFunction{DMprepare}{dm-prepare}
\SetKwFunction{DMcommit}{dm-commit}
\SetKwFunction{DMabort}{dm-abort}
\SetKwFunction{DMisprepare}{dm-is-true-prepare}
\SetKwFunction{DMiscommit}{dm-is-true-commit}
\SetKwFunction{Get}{get}
\SetKwFunction{Put}{put}
\SetKwFunction{Lock}{lock}
\SetKwFunction{Unlock}{unlock}
\SetKwFunction{Version}{version}
\SetKwFunction{NextVersion}{next-version}
\SetKwProg{Function}{upon}{}{}

% \Function{\TMbegin}{
%     $rset$ $\gets$ $wset$ $\gets$ $frset$ $\gets$ $fwset$ $\gets$ $cset$ $\gets$ $\emptyset$\;
% }{}
% \Function{\TMread{key}}{
%     $\Box$ $\gets$ \Future{key}\;
%     $frset$ $\gets$ $frset$ $\cup$ \{$\Box$\}\;
%     \Return{$\Box$}\;
% }{}
\Function{\TMread{$\bigtriangleup$}}{
    $key$ $\gets$ \Key{$\bigtriangleup$}\;
    % $\langle value$, $version\rangle$ $\gets$ \DMread{$key$}\;
    $\langle value$, $version\rangle$ $\gets$ \Get{$key$}\;
    $rset$ $\gets$ $rset$ $\cup$ \{$\langle key$, $version\rangle$\}\;
    \Return{\TMread{value}}\;
}{}
% \Function{\TMwrite{key, $\Box$}}{
%     $wset$ $\gets$ $wset$ $\cup$ \{$\langle key$, $\Box\rangle$\}\;
% }{}
% \Function{\TMwrite{$\bigtriangleup$, $\Box$}}{
%     $fwset$ $\gets$ $fwset$ $\cup$ \{$\langle\bigtriangleup$, $\Box\rangle$\}\;
% }{}
\Function{\TMis{$\Box$}}{
   $rvalues$ $\gets$ $\emptyset$\;
   \ForEach{$key$ $\in$ \Keys{$\Box$}}{
       % $rvalues$ $\gets$ $rvalues$ $\cup$ \{$\langle key$, \DMread{key}$\rangle$\}\;
       $rvalues$ $\gets$ $rvalues$ $\cup$ \{$\langle key$, \Get{key}$\rangle$\}\;
   }
   $result$ $\gets$ \Resolve{$\Box$, $rvalues$}\;
   $cset$ $\gets$ $cset$ $\cup$ \{$\langle\Box$, $result\rangle$\}\;
   \Return{result}\;
}{}
\Function{\TMcommit}{
    $rvalues$ $\gets$ $\emptyset$\;
    $result$ $\gets$ $\top$\;
    \lForEach{$\langle key$, $-\rangle$ $\in$ $wset$}{
        \Lock{$key$}
    }
    \ForEach{$\Box$ $\in$ $frset$}{
        $key$ $\gets$ \Key{$\Box$}\;
        \Lock{$key$}\;
        $rvalues$ $\gets$ $rvalues$ $\cup$ \{$\langle key$, \Get{$key$}$\rangle$\}\;
    }
    \ForEach{$\langle\bigtriangleup$, $\Box\rangle$ $\in$ $fwset$}{
        $key$ $\gets$ \Resolve{$\bigtriangleup$, $rvalues$}\;
        \Lock{$key$}\;
        $wset$ $\gets$ $wset$ $\cup$ \{$\langle key$, $\Box\rangle$\}\;
    }
    \ForEach{$\langle key$, $version\rangle$ $\in$ $rset$}{
        \lIf{$version$ $\neq$ \Version{$key$}}{
            $result$ $\gets$ $\bot$
        }
    }
    \ForEach{$\langle\Box$, $expected\rangle$ $\in$ $cset$}{
        $value$ $\gets$ \Resolve{$\Box$, $rvalues$}\;
        \lIf{value $\neq$ expected}{
            $result$ $\gets$ $\bot$
        }
    }
    \If{$result = \top$}{
        \ForEach{$\langle key$, $\Box\rangle$ $\in$ $wset$}{
            $value$ $\gets$ \Resolve{$\Box$, $rvalues$}\;
            $version$ $\gets$ \NextVersion{$key$}\; 
            \Put{$key$, $value$, $version$}\;
        }
    }
    \lForEach{$key \in wset \cup \Keys{$frset$}$}{
        \Unlock{$key$}
    }
    \Return{$\langle result$, $rvalues\rangle$}\;
}
	\caption{\api{lsd}-aware OCC protocol.}
	\label{fig:clt_occ_lsd}
\end{algorithm}
In a nutshell, OCC works as follows.
Each database object is associated with a version.
Reads record the object identity and the observed version in the \emph{read~set}.
Writes are buffered in the \emph{write~set} until the transaction attempts to commit, instead of modifying the database immediately.
Then, when a transaction attempts to commit, it atomically verifies if every object in the read~set is unchanged, i.e., if it is still in the same version that was read, and, if so, all buffered updates are applied, and the respective version numbers are incremented.
This atomic test and change is implemented in three steps:~(1)~lock the write~set,~(2)~validate the read~set, and (3)~perform the pending writes, if the validation was successful, and release the acquired locks.

Next, we describe the adaptations required for the remaining operations, as depicted in Figure~\ref{fig:clt_occ_lsd}.

\paragraph{\api{read(}$\bigtriangleup$\api{)}.}
Resolves the future-key~$\bigtriangleup$, i.e., compute its concrete value~$value$ and add the  observed version to the read~set, and then \api{read(}$value$\api{)}. (Returning a future.)

\paragraph{\api{is-true(}$\Box$\api{)}.}
Observes the current value of each key present in the condition~$\Box$, i.e., each future-value over which the condition is defined, resolve~$\Box$ using the observed values, add the result to the condition~set, and returns it.

\paragraph{\api{commit}.}
As we discussed, the commit protocol executes in three steps.
First, we lock the write~set and the future-write~set. However, the latter initially has its keys unresolved.
To resolve and then lock them, we first need to resolve the future~read~set because it contains the future-values of \api{read} operations that were delayed, and future-keys and future-values in the regular and future~write~sets are likely to depend on the future~read~set.
$\big($e.g., a transaction reads~$key$ and gets future-value~$\Box$, which it then uses to create a future ${\bigtriangleup = f(\Box)}$, according to some function~$f$, that the transaction uses as a future-key~---~\api{write(}$\bigtriangleup$\api{,...)}~---~and/or future-value~---~\api{write(...,}$\bigtriangleup$\api{)}.$\big)$
% After resolving the future~read~set, resolve the future~write~set's future-keys and lock them.
To guarantee that we resolve the future~read~set consistently, we first lock the respective keys.

In the second step, we validate the read~set.
In addition to the read~set, transactions also observe database state via conditions and the \api{is-true} operation, so we also validate each condition in the condition~set using the values obtained from the future~read~set.
In the final step, we resolve the buffered future-values, perform the writes, and release acquired locks.

To illustrate these steps, we will simulate the execution of our running example of Figure~\ref{subfig:api_lsd}.
First, the transaction issues the \api{read} operation for the item's stock.
This operation is local to the client, since it merely creates the future $\Box$ and returns it.
Then the transaction attempts to purchase $qty$ amount of items if there is enough stock. Let us assume that $qty = 10$.
Since the transaction does not know the concrete value of the item's stock, it uses the \api{is-true} operation to check whether there are at least $10$ items available.
Assume that, in this example execution, the transaction is operating on a database state where there are at least $10$ items in stock. Then, in order to maintain isolation, this condition must also hold when the transaction attempts to commit, and thus the transaction records the condition and its result in the condition~set for commit-time validation.
Finally, the transaction defines the necessary computation to update the stock value with the future $\bigtriangleup$, issues the \api{write} with it, and attempts to commit.
The \api{commit} operation will then atomically resolve the stock value $\Box$ to, for example, $42$, verify that $42 \ge 10$, and compute the new stock value $\bigtriangleup$ to be $42 - 10 = 32$.
Note that, when using standard OCC, any  concurrent write to the stock value causes the transaction to abort.
With \api{lsd}-aware OCC, instead, the transaction only aborts if between the time the \api{is-true} and \api{commit} operations are issued the stock value changes to a value below $10$.

\paragraph{Possible optimization.} % (fold
Since the \api{is-true} operations are validated at commit time to ensure isolation, it is possible to optimistically assume a specific result for an \api{is-true} operation \emph{without} communicating with the database.
Whether this behavior yields better performance or not depends on the success rate of the assumption:
if the assumption is correct we save one communication round with the database,
but if it is not, the transaction aborts, perhaps needlessly, and upon retry performs the \api{is-true} operation normally.
We evaluate this optimization in \S\ref{sec:evaluation}.
% noindent (end)
% subsubsection (end)
\subsubsection{2-phase locking (2PL)} % (fold)
\label{subsub:design_cc_2pl}
\begin{figure}[tb]
	\footnotesize
    \centering
    \begin{tabular}{r|c|c|c|c|c}
& R & W & R($p$) & W$\big(v : c(v)\big)$ & W$\big(v : \neg c(v)\big)$ \\
\midrule
R & \cellcolor{gray!25}$\checkmark$ & \cellcolor{gray!25}$-$ & $\checkmark$ & $-$ & $-$ \\
W & \cellcolor{gray!25}$-$ & \cellcolor{gray!25}$-$ & $-$ & $-$ & $-$ \\
R($p$) & $\checkmark$ & $-$ & $\checkmark$ & $\checkmark$ & $-$ \\
W$\big(v : c(v)\big)$ & $-$ & $-$ & $\checkmark$ & $-$ & $-$ \\
W$\big(v : \neg c(v)\big)$ & $-$ & $-$ & $-$ & $-$ & $-$ \\
    \end{tabular}
    \caption{
        2PL's lock compatibility matrix.
        The gray cells represent the standard 2PL matrix, and \api{lsd} introduces the remaining cells.
        A $\checkmark$ means that acquiring the lock in row mode succeeds when the lock is in column mode.
    }
    \label{fig:2pl_locks}
\end{figure}

\begin{algorithm}[htp]
\linespread{1.15}\selectfont
\small
\DontPrintSemicolon
\SetInd{0.1em}{0.5em}
\SetKwFunction{TMbegin}{begin}
\SetKwFunction{TMread}{read}
\SetKwFunction{TMis}{is-true}
\SetKwFunction{TMwrite}{write}
\SetKwFunction{TMcommit}{commit}
\SetKwFunction{TMabort}{abort}
\SetKwFunction{Future}{future}
\SetKwFunction{Keys}{keys}
\SetKwFunction{Key}{key}
\SetKwFunction{Resolve}{resolve}
\SetKwFunction{DMread}{dm-read}
\SetKwFunction{DMprepare}{dm-prepare}
\SetKwFunction{DMcommit}{dm-commit}
\SetKwFunction{DMabort}{dm-abort}
\SetKwFunction{DMisprepare}{dm-is-true-prepare}
\SetKwFunction{DMiscommit}{dm-is-true-commit}
\SetKwFunction{Get}{get}
\SetKwFunction{Put}{put}
\SetKwFunction{Lock}{lock}
\SetKwFunction{Unlock}{unlock}
\SetKwFunction{Lockpredicate}{add-condition}
\SetKwFunction{Unlockpredicate}{rem-condition}
\SetKwFunction{Lockcompatible}{lock-compatible}
\SetKwFunction{Version}{version}
\SetKwProg{Function}{upon}{}{}

% \Function{\TMbegin}{
%     $rset$ $\gets$ $wset$ $\gets$ $frset$ $\gets$ $fwset$ $\gets$ $cset$ $\gets$ $\emptyset$\;
% }{}
% \Function{\TMread{key}}{
%     $\Box$ $\gets$ \Future{key}\;
%     $frset$ $\gets$ $frset$ $\cup$ \{$\Box$\}\;
%     \Return{$\Box$}\;
% }{}
\Function{\TMread{$\bigtriangleup$}}{
    $key$ $\gets$ \Key{$\bigtriangleup$}\;
    % $value$ $\gets$ \DMread{$key$}\;
    \Lock{key}\;
    $value \gets $ \Get{key}\;
    $rset$ $\gets$ $rset$ $\cup$ \{$key$\}\;
    \Return{\TMread{value}}\;
}{}
% \Function{\TMwrite{key, $\Box$}}{
%     $wset$ $\gets$ $wset$ $\cup$ \{$\langle key$, $\Box\rangle$\}\;
% }{}
% \Function{\TMwrite{$\bigtriangleup$, $\Box$}}{
%     $fwset$ $\gets$ $fwset$ $\cup$ \{$\langle\bigtriangleup$, $\Box\rangle$\}\;
% }{}
\Function{\TMis{$\Box$}}{
    $rvalues$ $\gets$ $\emptyset$\;
    \ForEach{$key$ $\in$ \Keys{$\Box$}}{
        \Lock{$key$}\;
        $rvalues$ $\gets$ $rvalues$ $\cup$ \{\Get{$key$}\}\;
    }
    $result$ $\gets$ \Resolve{$\Box$, $rvalues$}\;
    \ForEach{$key$ $\in$ \Keys{$\Box$}}{
        \Lockpredicate{$key$, $\langle\Box$, $result\rangle$}\;
        \Unlock{$key$}\;
    }
    $cset$ $\gets$ $cset$ $\cup$ \{$\Box$\}\;
    \Return{$result$}\;
}{}
\Function{\TMcommit}{
    $rvalues$ $\gets$ $\emptyset$\;
    \ForEach{$\Box$ $\in$ $frset$}{
        $key$ $\gets$ \Key{$\Box$}\;
        \Lock{$key$}\;
        $rvalues$ $\gets$ $rvalues$ $\cup$ \{$\langle key$, \Get{$key$}$\rangle$\}\;
    }
    \ForEach{$\Box \in cset$}{
        \lForEach{$key$ $\in$ \Keys{$\Box$}}{
            \Unlockpredicate{$key$, $\Box$}
        }
    }
    $set$ $\gets$ $\emptyset$\;
    \ForEach{$\langle\bigtriangleup$, $\Box\rangle$ $\in$ $fwset$}{
        $key$ $\gets$ \Resolve{$\bigtriangleup$, $rvalues$}\;
        $set$ $\gets$ $set$ $\cup$ \{$\langle key$, $\Box\rangle$\}\;
    }
    $writes$ $\gets$ $\emptyset$\;
    \ForEach{$\langle key$, $\Box\rangle$ $\in$ $wset \cup set$}{
        $value$ $\gets$ \Resolve{$\Box$, $rvalues$}\;
        $writes$ $\gets$ $writes$ $\cup$ \{$\langle key$, $value\rangle$\}\;
    }
    \ForEach{$\langle key$, $value\rangle$ $\in$ $writes$}{
        \Lockcompatible{$key$, $writes$}\;
        \Put{$key$, $value$}\;
    }
    \lForEach{$key \in writes \cup rset \cup \Keys{frset}$}{
        \Unlock{$key$}
    }
    \Return{$\langle \top, rvalues \rangle$}\;
}{}
    \caption{\api{lsd}-aware 2PL protocol.}
    \label{fig:clt_2pl_lsd}
\end{algorithm}
2PL follows a rational opposite to OCC: instead of assuming that conflicts seldom happen, 2PL immediately acquires a lock when a transaction accesses an object to prevent conflicting transactions from breaking isolation.

The central idea of the adaptation of 2PL to \api{lsd}'s \api{is-true} operation is the novel concept of a \emph{condition~lock}, which is an extension of a \emph{read-write~lock}.
To understand the semantics of condition~locks, we first recall that read-write~locks can be acquired in either read or write mode (R or W).
The semantics of read-write~locks are then given by their compatibility matrix shown in gray in Figure~\ref{fig:2pl_locks}.
This shows that multiple readers, i.e., read-mode acquires, can proceed simultaneously, but writers are serialized.
Condition~locks, in turn, have two additional modes: \emph{read~condition} and \emph{write~value}.
The read~condition mode, R($c$), associates a condition~$c$ with the lock, signaling that a transaction has observed a value that respects the condition~$c$.
Other transactions can still successfully update a read~condition-locked object by acquiring the lock in write~value mode.
The write~value mode, W($v$), is aware of the value~$v$ that the transaction intends to assign to the object.
If the lock is in read~condition mode and the value~$v$ respects all the conditions that the lock holds, the write~mode acquire succeeds.
Otherwise it blocks as usual.
Note that the read condition mode is a generalization of the read mode: the latter is smilar to the former with a condition that always returns false regardless of the value other transactions intend to write.
% Using these additional modes the \api{lsd}-aware 2PL protocol is able to extract read-write concurrency that the standard 2PL protocol prevents.

Next, we describe the adaptations required for the remaining operations, which are also summarized in Figure~\ref{fig:clt_2pl_lsd}.

\paragraph{\api{read(}$\bigtriangleup$\api{)}.}
Resolves the future-key~$\bigtriangleup$, i.e., compute its concrete value~$value$ by locking~$key$, reading its value, and then \api{read(}$value$\api{)}. (Returning a future.)

\paragraph{\api{is-true(}$\Box$\api{)}.}
Atomically observes the current value of each key present in the condition~$\Box$ by locking all keys.
Resolve the condition~$\Box$ using the observed values, and downgrade the acquired locks to read~condition mode using $\Box$ and its result.

\paragraph{\api{commit}.}
Resolves the future~read~set by locking and performing the delayed reads.
Remove the conditions installed via the \api{is-true} operations since we already resolved the future~read~set.
Resolve all future-keys in the future~write~set, and all future-values in the future and concrete~write~set.
Given that we now know the transaction's full write~set, acquire the locks in write~value mode, perform the writes, and release the acquired locks.

Again, to better understand these steps, we will go through the steps of the execution of our running example of Figure~\ref{subfig:api_lsd}.
The transaction reads the item's stock, which is an operation local to the client.
Then the transaction attempts to purchase $qty$ amount of items if there is enough stock. Let us assume that $qty = 10$.
The transaction uses the \api{is-true} operation to check whether there are at least $10$ items available.
Again, assuming that this is the case, to maintain isolation this must be also true when the transaction commits.
To ensure this, a condition~lock is acquired, in read~condition mode, on the stock stating that its value must remain greater or equal to $10$.
The transaction proceeds to define the necessary computation to update the stock value with the future $\bigtriangleup$, issues the \api{write} with it, and attempts to commit.
The \api{commit} operation will then atomically resolve the stock value $\Box$ to, for example, $42$, remove the condition~${\Box \ge qty}$ from the stock lock, and compute the new stock value $\bigtriangleup$ to be $42 - 10 = 32$.
Then the transaction acquires the stock's lock in write~value mode with $32$, blocking only if there is any concurrent reader that installed a condition $c$ such that $\neg c(32)$.
(With standard 2PL \emph{any} concurrent reader would cause the transaction to block.)
Finally, the transaction modifies the stock to~$32$ and releases the locks.
\subsubsection{Multi-future conditions} % (fold)
\label{subsub:design_cc_predicates}
OCC and 2PL fundamentally differ on how they deal with the validity of conditions.
OCC does not ensure that a condition asserted via the \api{is-true} operation remains valid.
This is because write transactions are not aware of those conditions and can freely violate the conditions when they commit.
As such, it is up to a transaction that asserts a condition to validate it when the transaction attempt to commit to ensure isolation, i.e., the burden of dealing with conditions is on the readers.
In constrast, 2PL ensures that an asserted condition remains valid until the asserting transaction commits, as acquring a condition~lock in write~value mode will block if the value to be written violates any existing asserted condition, i.e., the burden of dealing with conditions is on the writers.

Dealing with conditions on the writer's side is more complex than on the reader's, and this complexity is exacerbated in the presence of conditions over more than one future. 
For example, consider two keys~$x$ and~$y$, with values $2$ and~$1$, respectively, read by some transaction~$t_1$ as futures~$\Box$~and~$\bigtriangleup$.
$t_1$ then executes \api{is-true(}$\{\Box > \bigtriangleup\}$\api{)}, which returns~$\top$~(because $2 > 1$).
Then assume that, concurrently, another transaction~$t_2$ attempts to write~$1$ to~$x$.
For~$t_2$ to acquire~$x$'s condition~lock in write~value~mode with value~$1$ and commit, the procedure to acquire the condition~lock in write~value mode~(\api{lock-compatible} in Figure~\ref{fig:clt_2pl_lsd}) can only grant the lock to~$t_1$ if~$1 > \bigtriangleup$ remains valid.
Thus, the locking~procedure must resolve~$\bigtriangleup$ to check the concrete validity of~$1 > 1$.
To do so, there are two possibilities. If~$t_2$ also reads~$y$, then it has acquired a read~lock on~$y$ so it can resolve~$\bigtriangleup$.
If not, the lock~procedure needs to resolve $\bigtriangleup$ in a way that ensures transactional isolation, e.g., acquiring a read~lock on~$y$ on behalf of~$t_2$.
% Multi-future conditions with 2PL become even more elaborate if the database is partitioned and the futures reside in different partitions; we elaborate in \S\ref{subsub:design_dtxns_2pl}.

Given the experience in the implementation of our prototype, we argue that the \api{is-true} operation is simpler to implement, and understand, using an optimistic approach.
Additionally, the experimental evaluation~(\S\ref{sec:evaluation}) using our prototype shows that the \api{lsd}-aware OCC protocol performs better than the \api{lsd}-aware 2PL protocol, so we conclude that future implementations of \api{lsd} should use OCC in most cases.
%{\bf [rodrigo: aqui pede uma conclusao destas duas ultimas frases. E entao? Vais excluir de todo? Ou ha situacoes em que ainda pode fazer sentido? Quais?]}
% subsubsection (end)
% subsection (end)
\subsection{Distributed transactions} % (fold)
\label{sub:design_dtxns}
% \begin{figure*}[tb]
% 	\centering
%     \begin{subfigure}[b]{0.49\textwidth}
% 		\centering
% 	    \includegraphics[width=0.9\linewidth]{diagrams/2pc-occ-trimmed.pdf}
%         \caption{2PC and OCC.}
%         \label{subfig:design_dtxns_occ}
%     \end{subfigure}
%     \begin{subfigure}[b]{0.49\textwidth}
% 		\centering
% 	    \includegraphics[width=0.9\linewidth]{diagrams/2pc-2pl-trimmed.pdf}
%         \caption{2PC and 2PL.}
%         \label{subfig:design_dtxns_2pl}
%     \end{subfigure}
%     \caption{
% 	    Distributed transactions using 2PC and (a)~OCC, and (b)~2PL.
% 	    C and P stand for the 2PC coordinator and participants, respectively.
% 	    The shaded area is the prepare~phase, and the subsequent area is the commit~phase.
% 	    The dashed actions are not required for all transactions.
% 	}
%     \label{fig:design_dtxns}
% \end{figure*}
So far we have discussed how to adapt both OCC and 2PL to exploit \api{lsd} in the context of a single server.
However, transactions may be distributed, i.e., span multiple servers, if the database is partitioned.
We now briefly sketch how to adapt 2-phase~commit~(2PC)~\cite{databasebook-1987}, the most widely used distributed commit protocol, to support \api{lsd}.
A more comprehensive discussion of \api{lsd} together with 2PC can be found in~\cite{phd_std:2018:tvale}

\api{lsd} introduces the future read~and write~sets, and condition~set.
The future write~set is of particular importance, since it depends on the future read~set.
This means that, in general, transactions that have a non-empty future write~set require an additional round of communication during 2PC's prepare~phase.
Each participant resolves, and returns, its portion of the future read~set in the regular communication round of the prepare~phase.
Armed with the resolved future read~set the coordinator can resolve the future write~set and send it to the required participants.

It is possible to circumvent the need for the additional communication round in the prepare~phase and send the future~write~set immediately in the first round if, for every entry in the future~write~set:~(1)~we can identify its future-key's partition without resolving it,~and~(2)~(all) the future(s) on which the future-key depends is (are) from the same partition it belongs to.
In our experiments we evaluate both cases: when \api{lsd} incurs in an additional communication round in 2PC, and when it does not.
% subsection (end)
% section (end)
\section{Evaluation} % (fold)
\label{sec:evaluation}
We implemented a partitioned, transactional, key-value store prototype, including all of the previously described design with the exception of multi-future conditions.
Each partition is implemented as a Thrift~\cite{thrift} non-blocking server, and data is stored in disk using RocksDB~\cite{rocksdb}.
Clients can execute transactions using the typical API~(\api{begin}, \api{read}, \api{write}, \api{commit}, and \api{abort} operations) or \api{lsd}'s API which features our proposed \api{read}, \api{write}, and \api{is-true} operations.
We implemented both classical OCC and 2PL, and also both their \api{lsd}-aware variants for \api{lsd} transactions.
Distributed transactions commit using 2PC.
We resolve deadlocks that may arise in 2PL or 2PC using the wound-wait strategy.

We conducted an experimental evaluation of our \api{lsd} prototype on a private gigabit ethernet cluster.
Each server runs on a machine with a 2Ghz Intel Xeon E5-2620 processor, 32GB of RAM, and a 7200 RPM hard drive.
Clients run on the various remaining machines with AMD and Intel processors, and communicate with the servers using Thrift RPCs.

Each data point reports the average of 5 runs.
% In all of our experiments, the standard deviation across runs was below Y\%.
Our evaluation seeks to answer the following questions:
\begin{itemize}
	\item Does \api{lsd} improve the performance of realistic applications under contention?~(\S\ref{subsub:evaluation_tpcc_high})
	\item What is \api{lsd}'s overhead when contention is low?~(\S\ref{subsub:evaluation_tpcc_low})
	\item How do \api{lsd}'s benefits vary across various deployment scenarios, such as with a single database, or with a partitioned database and distributed transactions?~(\S\ref{sub:evaluation_tpcc})
	\item What is the impact of an increasing amount of contention with and without conditions?~(\S\ref{sub:evaluation_micro})
	% \item What is the impact of increasing the number of write~functions that transactions perform?
	% \item What is the impact of increasing the number of predicates that transactions assert?
\end{itemize}
\subsection{Realistic application: TPC-C} % (fold)
\label{sub:evaluation_tpcc}
\begin{figure*}[tb]
% high contention
	% 1 server
	\begin{subfigure}[t]{0.33\textwidth}
		\centering
		\begin{tikzpicture}
			\begin{axis}[
				xlabel={Throughput (commits/s)},
				xlabel near ticks,
				x label style={align=center,font=\small},
				x tick label style={font=\small},
				ylabel={Avg. latency (ms)},
				ylabel near ticks,
				y label style={align=center,font=\small},
				y tick label style={font=\small},
				xmin=0, xtick={0, 250, 500, 750, 1000, 1250}, xmax=1250,
				xticklabels={0, 250, 500, 750, 1K, 1.25K},
				ymin=0, ytick={0, 40, 80, 120, 160, 200}, ymax=200,
				ymajorgrids=true, yminorgrids=true, 
				xmajorgrids=true, xminorgrids=true,
				grid style={dotted,gray!50},
				axis on top,
				legend cell align={left},
				legend entries={{OCC}},
				legend columns=1,
				legend style={fill=none,draw=none,font=\footnotesize},
				width=\linewidth,
				height=0.6\linewidth
			]
				\addplot [mark size=2, thick, mark options={solid}, dashed, red!100!black, mark=o] table {
309.333333333333 19.2842145211667
% 362.333333333333 32.8970125961667
362.666666666667 48.0872838833333
% 308.666666666667 57.1732823515667
% 272.333333333333 71.8958132144
% 239.666666666667 84.8318985728667
225 92.5760406932333
% 238.333333333333 94.1296212824
% 220.666666666667 105.418483348667
% 206.666666666667 116.530840730333
% 195.333333333333 128.496510190333
% 188.333333333333 134.885079127667
153.333333333333 165.634289947333
% 152.333333333333 164.517289072667
% 139.333333333333 179.919796713
121.666666666667 182.043825775667
				};
				\addplot [mark size=2, thick, green!50!black, mark=*] table {
349.333333333333 17.0692830847333
% 554 21.5086911287667
713.666666666667 25.0930780364
% 776.333333333333 33.3311301745333
% 822 36.3267338103
% 847 39.9957102203333
838 45.2110256808667
% 845.333333333333 49.5275931786667
% 895.333333333333 50.8071440663667
% 955.666666666667 52.2953804121333
% 939.666666666667 57.2564564579333
% 974.666666666667 59.4188976687667
1024.33333333333 68.1561000335667
% 1027.66666666667 75.7174572368333
% 982 87.4444226574667
970.333333333333 96.7767735535
				};
				\addplot [mark size=2, thick, mark options={solid}, dashed, red!100!black, mark=triangle] table {
306.666666666667 19.4666070009333
% 372.333333333333 32.1101065277333
441 40.5731058557
% 346 74.0180539957
% 360.666666666667 80.1110584073667
% 348 89.1220720013
323.666666666667 94.1614087094667
% 326.666666666667 99.2233000595
% 303.333333333333 113.581665629
% 321.333333333333 113.2674039701
% 320 119.746231189
% 336 125.495742665667
341 138.29844486
% 333.333333333333 149.769616639333
% 310.666666666667 168.996391727333
312.333333333333 175.404986737
				};
				\addplot [mark size=2, thick, green!50!black, mark=triangle*] table {
236.666666666667 25.2050203984333
% 379.333333333333 31.4769873195667
520.333333333333 34.4686194439
% 604.333333333333 42.9051687377
% 636 46.9992692897
% 668 50.7310143364
677.666666666667 55.8234971398667
% 668 62.7095984015667
% 673.333333333333 68.1235578648
% 668.666666666667 74.6029691685
% 713.333333333333 75.5560346144667
% 765 75.6624161396333
834.666666666667 83.5988501077333
% 796.666666666667 97.9164499188333
% 803 107.017860865
839 111.629468784667
				};
			\end{axis}
		\end{tikzpicture}
		\caption{$1$~server, high contention.}
		\label{subfig:tpcc_high_s1}
	\end{subfigure}
	% 3 servers, warehouse
	\begin{subfigure}[t]{0.33\textwidth}
		\centering
		\begin{tikzpicture}
			\begin{axis}[
				xlabel={Throughput (commits/s)},
				xlabel near ticks,
				x label style={align=center,font=\small},
				x tick label style={font=\small},
				ylabel={Avg. latency (ms)},
				ylabel near ticks,
				y label style={align=center,font=\small},
				y tick label style={font=\small},
				xmin=0, xtick={0, 400, 800, 1200, 1600, 2000}, xmax=2000,
				xticklabels={0, 400, 800, 1.2K, 1.6K, 2K},
				ymin=0, ytick={0, 20, 40, 60, 80, 100}, ymax=100,
				ymajorgrids=true, yminorgrids=true, 
				xmajorgrids=true, xminorgrids=true,
				grid style={dotted,gray!50},
				axis on top,
				legend cell align={left},
				legend entries={,{OCC-LSD}},
				legend columns=1,
				legend style={fill=none,draw=none,font=\footnotesize},
				width=\linewidth,
				height=0.6\linewidth
			]
				\addplot [mark size=2, thick, mark options={solid}, dashed, red!100!black, mark=o] table {
275.333333333333 21.629841579
% 464.666666666667 25.6768355374333
599.666666666667 29.8888181375
% 645 39.6507131099333
% 681 42.946056709
% 706 46.9508285893
720.333333333333 50.2991524694667
% 733.333333333333 53.7933402655667
% 722.333333333333 58.6822208967
% 713.666666666667 62.2118970801333
% 679.666666666667 64.5285839117
% 626.666666666667 64.5849342800333
% 645.333333333333 69.9283530329667
567.666666666667 71.2059327611333
% 474 72.3329397568
386.666666666667 76.6222245391
403.333333333333 75.6756234464
				};
				\addplot [mark size=2, thick, green!50!black, mark=*] table {
264.666666666667 22.4907851521
% 511 23.3247225445333
739.666666666667 24.1945700626
% 880.666666666667 29.3644559567667
% 991.666666666667 30.0989328298
% 1094 30.8404183218
1189.66666666667 31.7628687300667
% 1278 32.5491705502667
% 1358.33333333333 33.6994295416
% 1452 34.0210841711667
% 1537.33333333333 34.7843096107667
% 1616 35.3281313464
% 1803 38.6566173877333
1837 42.2908088341333
% 1857.66666666667 45.0978603201
1934.66666666667 48.4223022498
1915.66666666667 53.2055727472
				};
				\addplot [mark size=2, thick, mark options={solid}, dashed, red!100!black, mark=triangle] table {
276 21.5907913335333
% 481.666666666667 24.7644170827667
646.666666666667 27.7090345273333
% 653 39.6668637506333
% 685 43.6038007918667
% 716.666666666667 47.1073149647667
760 49.7624964419333
% 789.666666666667 52.9083263243
% 814 56.2122930480667
% 845 58.7578737109333
% 865.333333333333 61.9598278731
% 879.666666666667 64.909430912
% 944.666666666667 72.4839066853
906.333333333333 80.4829641281667
% 902 88.8424526469667
859.333333333333 96.4831793011333
858 99.2852112797
				};
				\addplot [mark size=2, thick, green!50!black, mark=triangle*] table {
193 30.9475134683333
% 364.666666666667 32.7689956045667
528.666666666667 33.8793973410667
% 633.333333333333 40.8086717096333
% 711.666666666667 41.9824762172667
% 795.333333333333 42.5992578756333
850 44.5204582742667
% 953.333333333333 43.8182931559
% 1014.66666666667 45.1743482980333
% 1080.66666666667 46.1025604707333
% 1151.33333333333 46.767579702
% 1223.66666666667 47.2151350944667
% 1334.33333333333 52.3034992240667
1410.66666666667 55.0661701688
% 1391.66666666667 60.1824247521333
1455 64.4034929147333
1519.66666666667 70.6227259852
				};
			\end{axis}
		\end{tikzpicture}
		\caption{\footnotesize $3$~servers, high contention, part. by warehouse.}
		\label{subfig:tpcc_high_s3_tpcc}
	\end{subfigure}
	% 3 servers, hash
	\begin{subfigure}[t]{0.33\textwidth}
		\centering
		\begin{tikzpicture}
			\begin{axis}[
				xlabel={Throughput (commits/s)},
				xlabel near ticks,
				x label style={align=center,font=\small},
				x tick label style={font=\small},
				ylabel={Avg. latency (ms)},
				ylabel near ticks,
				y label style={align=center,font=\small},
				y tick label style={font=\small},
				xmin=0, xtick={0, 150, 300, 450, 600, 750}, xmax=750,
				ymin=0, ytick={0, 40, 80, 120, 160, 200}, ymax=200,
				ymajorgrids=true, yminorgrids=true, 
				xmajorgrids=true, xminorgrids=true,
				grid style={dotted,gray!50},
				axis on top,
				legend cell align={left},
				legend entries={},
				legend columns=1,
				legend style={fill=none,draw=none,font=\footnotesize},
				width=\linewidth,
				height=0.6\linewidth
			]
				\addplot [mark size=2, thick, mark options={solid}, dashed, red!100!black, mark=o] table {
187.666666666667 31.7187764334667
% 270.666666666667 44.0861474226667
287 58.4519621702667
% 271.333333333333 65.4466085306667
% 260.666666666667 71.6456892690667
% 262.333333333333 81.8441020964667
251.666666666667 89.4201928262667
% 246.333333333333 96.9819669917667
% 240.666666666667 104.277369530333
% 228 111.582841948333
% 232.333333333333 119.500465652
% 216 126.228468583667
% 190 139.863538193333
183 149.114080357333
% 174.666666666667 140.868289671667
% 168.333333333333 141.831972294333
141.666666666667 161.749466254333
				};
				\addplot [mark size=2, thick, green!50!black, mark=*] table {
149 39.9954699767333
% 269.333333333333 44.3947575853333
362.333333333333 49.397304622
% 401 63.6486397464333
% 426 68.6593347974
% 462.333333333333 71.6669122533333
472 77.0192427326
% 491.333333333333 80.5803927567333
% 516.333333333333 84.5156968423
% 551 83.9460596040667
% 545 89.9997067501667
% 567.666666666667 90.3622306352333
% 548.333333333333 105.117431075667
522.666666666667 115.291564877667
% 537.666666666667 121.566544739667
% 506.333333333333 130.102025188
507 135.594595129
				};
				\addplot [mark size=2, thick, mark options={solid}, dashed, red!100!black, mark=triangle] table {
199 29.9871281804333
% 304.666666666667 39.1509264674667
356.666666666667 50.285696982
% 286.333333333333 86.7479092357667
% 281 95.4868614047667
% 266.666666666667 103.864340772667
266.333333333333 106.103504171667
% 263.333333333333 117.241380084667
% 249 127.820418739667
% 255 132.764746115667
% 271 136.475449050667
% 264.333333333333 143.110351499
% 263.666666666667 150.180012676667
266.333333333333 164.688131906667
% 262 176.253719701667
% 257.333333333333 182.856990218
239 190.561409818667
				};
				\addplot [mark size=2, thick, green!50!black, mark=triangle*] table {
125.666666666667 47.4713828301
% 223.333333333333 53.6027616415667
308 58.1794155798667
% 338.333333333333 74.0670193034667
% 371.333333333333 76.6168558501
% 404.333333333333 79.1670861364333
425 83.7846666738333
% 452 85.9362660822333
% 470 89.5528344738667
% 479 93.3035106028333
% 497.666666666667 95.4182130454333
% 523.333333333333 98.1538178785
% 522 110.908508503
504 118.884847515667
% 493 124.643330962667
% 499.333333333333 132.267400335333
484.666666666667 144.961654288333
				};
			\end{axis}
		\end{tikzpicture}
		\caption{$3$~servers, high contention, part. by hash.}
		\label{subfig:tpcc_high_s3_hash}
	\end{subfigure}
% low contention
	% 1 server
	\begin{subfigure}[t]{0.33\textwidth}
		\centering
		\begin{tikzpicture}
			\begin{axis}[
				xlabel={Throughput (commits/s)},
				xlabel near ticks,
				x label style={align=center,font=\small},
				x tick label style={font=\small},
				ylabel={Avg. latency (ms)},
				ylabel near ticks,
				y label style={align=center,font=\small},
				y tick label style={font=\small},
				xmin=0, xtick={0, 200, 400, 600, 800, 1000}, xmax=1000,
				xticklabels={0, 200, 400, 600, 800, 1K},
				ymin=0, ytick={0, 30, 60, 90, 120, 150}, ymax=150,
				ymajorgrids=true, yminorgrids=true, 
				xmajorgrids=true, xminorgrids=true,
				grid style={dotted,gray!50},
				axis on top,
				legend cell align={left},
				legend entries={,,{2PL}},
				legend columns=1,
				legend pos=north west,
				legend style={fill=none,draw=none,font=\footnotesize},
				width=\linewidth,
				height=0.6\linewidth
			]
				\addplot [mark size=2, thick, mark options={solid}, dashed, red!100!black, mark=o] table {
368 16.1949506161333
% 524 22.7789918125
647.666666666667 27.6955433681667
% 681 37.9942465252
% 681.666666666667 43.9321055196667
% 721 47.070617573
715.666666666667 52.8730619629
% 731.666666666667 57.3872278305333
% 714.333333333333 64.200919683
% 737 67.7804351602
% 822.666666666667 64.5843314332
% 782.666666666667 73.6629653248667
792.666666666667 87.6408090807667
% 740.333333333333 104.702149795433
% 739.333333333333 114.751752807
691 132.597655604333
				};
				\addplot [mark size=2, thick, green!50!black, mark=*] table {
334.666666666667 17.8136288256333
% 490.666666666667 24.3373417432333
629.666666666667 28.4647529352333
% 704.666666666667 36.7499281515667
% 737.333333333333 40.5475331403667
% 728.666666666667 46.5520641241
758.666666666667 50.0825255044
% 774 54.5426153083333
% 743.666666666667 61.6289021641333
% 802 62.3824410382333
% 845.333333333333 63.8313379482333
% 808.666666666667 71.4631634968667
801.333333333333 87.203227489
% 872.666666666667 89.5089346609333
% 854.333333333333 100.682225931667
819 114.452190726
				};
				\addplot [mark size=2, thick, mark options={solid}, dashed, red!100!black, mark=triangle] table {
367.666666666667 16.2095960966667
% 521.333333333333 22.9070789466667
649.333333333333 27.6023638613667
% 661.333333333333 39.1581397334667
% 649.666666666667 46.0207217496
% 676.666666666667 50.0514876598667
664.333333333333 56.4674590249333
% 690.666666666667 60.5611817349667
% 784.333333333333 58.4172391605
% 754.666666666667 65.3121510365333
% 781 68.9938495874333
% 795.333333333333 72.8845619646
798.333333333333 87.6327312023667
% 764.666666666667 101.411790262167
% 796.333333333333 107.279125893667
800.666666666667 116.036374653333
				};
				\addplot [mark size=2, thick, green!50!black, mark=triangle*] table {
228 26.1550489100667
% 360.333333333333 33.1680487177
480.333333333333 37.3207549191
% 538 48.1487869959333
% 576 51.9317613620667
% 587 57.7317551338
576 65.7432888614667
% 567.666666666667 73.9262411861667
% 660.333333333333 69.8308922955
% 618 80.7062981849333
% 630.333333333333 85.4794928805667
% 657.333333333333 88.1271521803333
654 106.572030063333
% 664.666666666667 117.319378656333
% 681 126.259598528
698.333333333333 134.515108749333
				};
			\end{axis}
		\end{tikzpicture}
		\caption{$1$~server, low contention.}
		\label{subfig:tpcc_low_s1}
	\end{subfigure}
	% 3 servers, warehouse
	\begin{subfigure}[t]{0.33\textwidth}
		\centering
		\begin{tikzpicture}
			\begin{axis}[
				xlabel={Throughput (commits/s)},
				xlabel near ticks,
				x label style={align=center,font=\small},
				x tick label style={font=\small},
				ylabel={Avg. latency (ms)},
				ylabel near ticks,
				y label style={align=center,font=\small},
				y tick label style={font=\small},
				xmin=0, xtick={0, 400, 800, 1200, 1600, 2000}, xmax=2000,
				xticklabels={0, 400, 800, 1.2K, 1.6K, 2K},
				ymin=0, ytick={0, 15, 30, 45, 60, 75}, ymax=75,
				ymajorgrids=true, yminorgrids=true, 
				xmajorgrids=true, xminorgrids=true,
				grid style={dotted,gray!50},
				axis on top,
				legend cell align={left},
				legend entries={,,,{2PL-LSD}},
				legend columns=1,
				legend pos=north west,
				legend style={fill=none,draw=none,font=\footnotesize},
				width=\linewidth,
				height=0.6\linewidth
			]
				\addplot [mark size=2, thick, mark options={solid}, dashed, red!100!black, mark=o] table {
299 19.9275564904333
% 546.666666666667 21.8255022421
757.666666666667 23.6207668823333
% 916.666666666667 28.2042588935333
% 1011.33333333333 29.5029099755333
% 1085 31.1816327693
1181.33333333333 32.0215875172667
% 1257.66666666667 33.2217804506
% 1317.33333333333 34.6106716135667
% 1394 35.6379557266
% 1470 36.5618591628667
% 1529 37.8004276008
% 1618.33333333333 43.0823091325667
1664 45.8901201081333
% 1629.66666666667 50.5965009065
1662.66666666667 54.9796085231333
1656.66666666667 60.8652694292
				};
				\addplot [mark size=2, thick, green!50!black, mark=*] table {
265.666666666667 22.418497774
% 513.666666666667 23.2165834661333
731 24.4858454161333
% 869.666666666667 29.7451922193667
% 981.666666666667 30.3974969511667
% 1085 31.1692787884333
1177.66666666667 32.1039433029
% 1262 32.8666436191667
% 1339.66666666667 33.9809529723667
% 1419.33333333333 34.7915729845333
% 1506 35.6940252131333
% 1562.66666666667 36.5221639824667
% 1755.66666666667 39.7194745423667
1774 43.0231738215
% 1809.33333333333 46.0142182926333
1873.66666666667 49.7817276805333
1891.33333333333 53.9004899537
				};
				\addplot [mark size=2, thick, mark options={solid}, dashed, red!100!black, mark=triangle] table {
291.333333333333 20.435444898
% 548.333333333333 21.7488803511333
771 23.2144311765667
% 900 28.7437338742333
% 1008 29.6241534416667
% 1108.33333333333 30.50245026
1181 32.0113488605667
% 1261.66666666667 33.1298850178667
% 1338.66666666667 34.1912304476667
% 1378.33333333333 35.6729444407667
% 1466.66666666667 36.6538364967667
% 1517.33333333333 37.9010592483667
% 1688.66666666667 41.2801295248
1648 45.4671740927333
% 1646.33333333333 49.6679934076
1749 53.5421295104667
1786.33333333333 59.3730834006333
				};
				\addplot [mark size=2, thick, green!50!black, mark=triangle*] table {
189 31.5748596229667
% 359.666666666667 33.2166725419
517.333333333333 34.6319601189333
% 620.666666666667 41.7061550523
% 704.666666666667 42.3970915321
% 783.666666666667 43.2657927228667
859.333333333333 44.0222506075
% 933 44.8112224373667
% 1013.66666666667 45.2067934680333
% 1077.66666666667 46.2423578048
% 1118.66666666667 47.8378481302
% 1185.33333333333 48.7889357035
% 1316 53.0064639264333
1360.66666666667 56.7833346198
% 1382 61.4661990146333
1442 65.0008586342333
1476 71.0160547327333
				};
			\end{axis}
		\end{tikzpicture}
		\caption{\footnotesize $3$~servers, low contention, part. by warehouse.}
		\label{subfig:tpcc_low_s3_tpcc}
	\end{subfigure}
	% 3 servers, hash
	\begin{subfigure}[t]{0.33\textwidth}
		\centering
		\begin{tikzpicture}
			\begin{axis}[
				xlabel={Throughput (commits/s)},
				xlabel near ticks,
				x label style={align=center,font=\small},
				x tick label style={font=\small},
				ylabel={Avg. latency (ms)},
				ylabel near ticks,
				y label style={align=center,font=\small},
				y tick label style={font=\small},
				xmin=0, xtick={0, 250, 500, 750, 1000, 1250}, xmax=1250,
				xticklabels={0, 250, 500, 750, 1K, 1.25K},
				ymin=0, ytick={0, 25, 50, 75, 100, 125}, ymax=125,
				ymajorgrids=true, yminorgrids=true, 
				xmajorgrids=true, xminorgrids=true,
				grid style={dotted,gray!50},
				axis on top,
				legend cell align={left},
				legend entries={},
				legend columns=1,
				legend pos=north west,
				legend style={fill=none,draw=none,font=\footnotesize},
				width=\linewidth,
				height=0.6\linewidth
			]
				\addplot [mark size=2, thick, mark options={solid}, dashed, red!100!black, mark=o] table {
241.666666666667 24.6794035843
% 453.333333333333 26.3264928559
619.333333333333 28.8965353932
% 726.666666666667 35.5864372894667
% 809.666666666667 36.8921970750667
% 871.333333333333 38.8291286669667
918.666666666667 41.1948754624333
% 943.666666666667 43.2687535474667
% 1006.66666666667 45.5121227422333
% 1027.33333333333 48.1381855217
% 1074.66666666667 49.9758683618
% 1073 52.3446813229333
% 1113.66666666667 59.3045667421333
1145 66.5049383824667
% 1122.33333333333 72.3155838058
1105.33333333333 77.997454778
1165 92.4912507686
				};
				\addplot [mark size=2, thick, green!50!black, mark=*] table {
159.333333333333 37.3850200231333
% 314 38.0503656201
446 40.2131323570667
% 536 48.3254745746667
% 603.666666666667 49.5245886870667
% 673.666666666667 50.2641626070667
731.333333333333 51.5930324715
% 794.666666666667 52.6626434716
% 853.666666666667 53.7091143968
% 911.333333333333 54.7276254128
% 959.333333333333 56.0928384817667
% 1008 57.2520126756333
% 1034.66666666667 64.5616615836667
1053.33333333333 70.1979945641
% 1120.33333333333 76.2459208685667
1159.33333333333 80.4875682429333
1208.66666666667 89.0720310541333
				};
				\addplot [mark size=2, thick, mark options={solid}, dashed, red!100!black, mark=triangle] table {
239 24.9133873612
% 451.333333333333 26.4421445405
617.333333333333 28.9866058210667
% 740.666666666667 34.9223007643667
% 808.666666666667 36.9177949503667
% 879.666666666667 38.4661141983
929.333333333333 40.7055369608333
% 979.666666666667 42.6880239209
% 1016 45.0959688290333
% 1045 47.1507179666667
% 1097.66666666667 48.7505423647
% 1107.33333333333 51.5737211006
% 1135 58.7081374013667
1202 63.7510686146667
% 1168.66666666667 69.7366816323333
1191.66666666667 75.3468548208333
1193 84.4684207127333
				};
				\addplot [mark size=2, thick, green!50!black, mark=triangle*] table {
136.333333333333 43.8393538779
% 247.333333333333 48.3630557648667
363.666666666667 49.3419327378667
% 436.666666666667 59.3934035840333
% 486.666666666667 61.3943749479
% 545.333333333333 62.1295341292333
602.666666666667 62.7539765573
% 632 64.7099403228667
% 701 65.1525718211333
% 745.333333333333 66.9022905562667
% 796.666666666667 67.4928058919333
% 823.333333333333 70.2283970142333
% 874.333333333333 78.3603868102667
917.666666666667 83.8544010351333
% 938.333333333333 90.3257457902
970.666666666667 96.2092412428
951.666666666667 105.139146766667
				};
			\end{axis}
		\end{tikzpicture}
		\caption{$3$~servers, low contention, part. by hash.}
		\label{subfig:tpcc_low_s3_hash}
	\end{subfigure}
	\caption{
		Performance of TPC-C on a workload using:
		1~server with high~(a) and low~(b) contention;
		3~servers with partitioning by warehouse~(b,e); and
		3~servers with partitioning by hash~(c,f).
	}
\end{figure*}
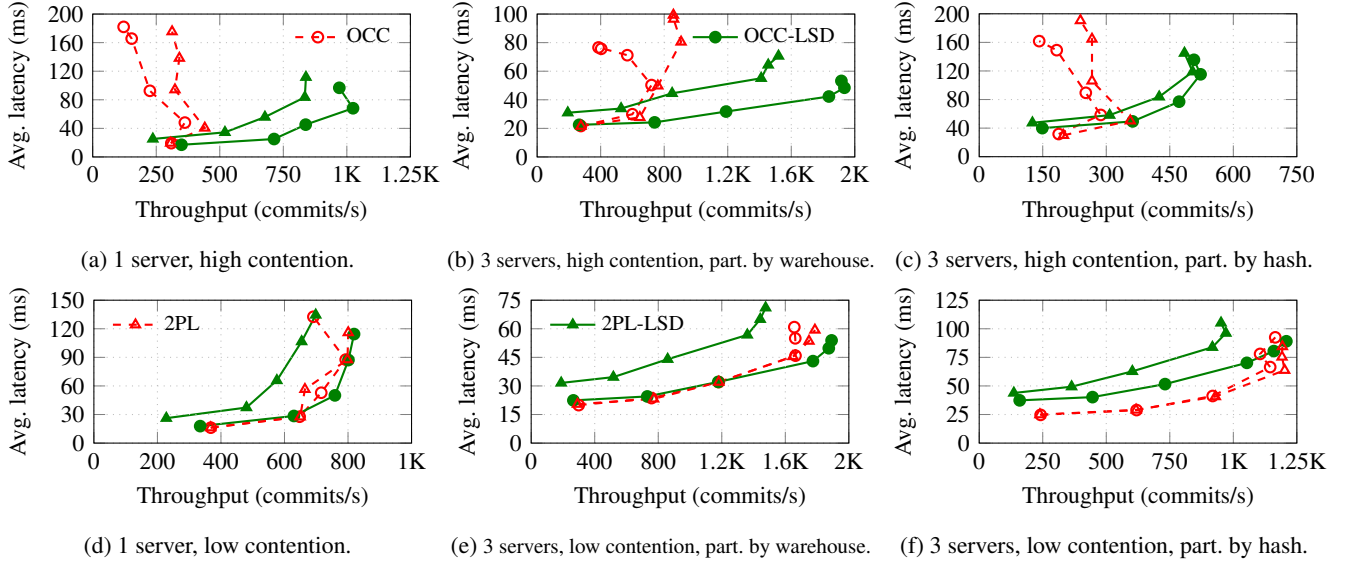
We used the popular TPC-C benchmark~\cite{tpcc} to assess \api{lsd}'s ability to improve performance of realistic applications under contention, as well as its overhead, on different deployment scenarios.
\api{lsd} was particularly helpful for the two core transactions of the workload: Payment and New~Order.
For example, both make use of write functions to modify client balance and stock values, and the latter also uses conditions.

We experimented with TPC-C under three different deployments:~(a)~a centralized database,~(b)~a partitioned database using an application-specific partitioning policy, and~(c)~a partitioned database using an application-agnostic partitioning policy.
We executed TPC-C with a high and low contention workload in each deployment.

\paragraph{Setup.}
We setup each deployment as follows.
The centralized data\-base~(a) uses a single server that stores the entire data.
The database partitioned using an application-specific policy~(b) uses $3$~servers.
The data associated with a particular warehouse is stored within a single server.
The remaining data, such as item information, is partitioned across all servers via hashing.
Finally, the database partitioned using an application-agnostic policy~(c) also uses~$3$ servers.
Data is partitioned across all servers via hashing.
\subsubsection{High contention} % (fold)

\label{subsub:evaluation_tpcc_high}
In TPC-C, the level of contention is proportional to the number of warehouses, so we loaded the database with the minimum number of warehouses applicable to each deployment~(as detailed below) and then executed TPC-C with an increasing number of clients.
Figures~\ref{subfig:tpcc_high_s1},~\ref{subfig:tpcc_high_s3_tpcc}, and~\ref{subfig:tpcc_high_s3_hash}, compare the throughput, measured in committed transactions per second~(x~axis), and the corresponding average transaction execution latency, measured in in milliseconds~(y~axis), of OCC and 2PL with and without~\api{lsd}.

\paragraph{Centralized deployment (Figure~\ref{subfig:tpcc_high_s1}).}% (fold
We loaded the database with $1$~warehouse.
The \api{lsd}-aware OCC variant achieved a peak throughput of ${\approx1}$K~committed transactions per second with an average latency of ${\approx70}$~ms, which amounts to ${\approx6.5\times}$ higher throughput and ${\approx2.5\times}$ lower latency than standard OCC under the same load.
The \api{lsd}-aware 2PL variant achieved a peak throughput of ${\approx850}$~committed transactions per second with an average latency of ${\approx80}$~ms, which amounts to ${\approx2.5\times}$ higher throughput and ${\approx1.5\times}$ lower latency than standard 2PL under the same load.
% noindent (end)

\paragraph{Partitioned deployment using application-specific policy (Figure~\ref{subfig:tpcc_high_s3_tpcc}).} % (fold)
We loaded the database with $3$~warehouses.
Data was partitioned across the servers by warehouse, i.e., each server hosts a single warehouse.
This scenario allows for the presence of distributed transactions. 
Distributed \api{lsd} transactions commit using the regular 2PC protocol, i.e., without incurring in the additional communication rounds discussed in \S\ref{sub:design_dtxns}, thanks to the application-specific partitioning policy.
The \api{lsd}-aware OCC variant achieved a peak throughput of $\approx2$K~committed transactions per second with an average latency of $\approx50$~ms, which amounts to $\approx5\times$ higher throughput and $\approx1.5\times$ lower latency than standard OCC under the same load.
The \api{lsd}-aware 2PL variant achieved a peak throughput of $\approx1.5$K~committed transactions per second with an average latency of $\approx60$~ms, which amounts to $\approx1.5\times$ higher throughput and $\approx1.3\times$ lower latency than standard 2PL under the same load.
% noindent (end)

\paragraph{Partitioned deployment using application-agnostic policy (Figure~\ref{subfig:tpcc_high_s3_hash}).} % (fold)
We loaded the database with a single warehouse, and all data is partitioned across the servers using hashing.
By using an application-agnostic partitioning policy, such as hashing, distributed \api{lsd} transactions may need an additional communication round to commit using 2PC.
This is the case for the New-Order transaction, which comprises almost half of the workload.
Despite the additional communication round, the \api{lsd}-aware OCC variant achieved a peak throughput of $\approx500$~committed transactions per second with an average latency of $\approx120$~ms, which amounts to $\approx2.8\times$ higher throughput and $\approx1.3\times$ lower latency than standard OCC under the same load.
The \api{lsd}-aware 2PL variant achieved a peak throughput of $\approx500$~committed transactions per second with an average latency of $\approx120$~ms, which amounts to $\approx1.8\times$ higher throughput and $\approx1.3\times$ lower latency than standard 2PL under the same load.
% noindent (end)

\paragraph{Discussion.} % (fold
This workload highlights the benefits of \api{lsd}.
For example, under the standard interface semantics, any two concurrent New-Order transactions conflict if:~(a)~they operate on the same district~(conflicting accesses to the district's order identifier counter), or~(b)~they order the same item~(conflicting accesses to the item's stock).
Under OCC only one of the concurrent transactions commits and the other aborts.
Under 2PL one of the transactions queues behind the other when it attempts to acquire the lock held by the other.
In both cases one of the transactions prevents the other from executing, leading to an effective serialization of their execution.
With \api{lsd}, New-Order transactions delay their accesses to the district's order identifier counter until commit time, so these accesses do \emph{not} result in aborts under OCC, nor queueing during transaction execution under 2PL.
Furthermore, any two New-Order transactions that order the same item only conflict if both attempt to buy the entire remaining stock.
\api{lsd}'s benefits translate in practice to higher throughput and lower latency under contention due to less aborts~(blocks) under OCC~(2PL).
For example, in the data point where \api{lsd} transactions achieve their peak throughput on Figure~\ref{subfig:tpcc_high_s1}, $\approx92\%$~of OCC transactions abort, whereas this number drops to $\approx8\%$ with the \api{lsd}-aware variant. 

It is worth noting that our \api{lsd}-aware 2PL implementation incurs in higher overhead than its OCC counterpart.
While there still may be room for optimization of our prototype, the \api{lsd}-aware 2PL has fundamentally more overhead than its OCC counterpart because condition~locks are a more complex technique than condition~validation.
The combination of this higher overhead of \api{lsd}-aware 2PL with the fact that unlike the usual OCC implementations, our \api{LSD}-aware variant of OCC presents low abort rate under high contention, leads to to a somewhat surprising result: the \api{LSD}-aware variant of OCC performs better than its 2PL counterpart under high contention.
% noindent (end)
% subsubsection (end)
\subsubsection{Low contention} % (fold)
\label{subsub:evaluation_tpcc_low}
In the previous section, we evaluated \api{lsd} using a TPC-C workload with high contention, which is the type of workload that \api{lsd} benefits.
In this section we describe our evaluation of \api{lsd} in the opposite scenario: a TPC-C workload with low contention.
Specifically, we increased the number of warehouses in the workload from~$1$~to~$32$.
% Figure~\ref{fig:tpcc_low} reports the results we obtained.

In both the centralized~(Figure~\ref{subfig:tpcc_low_s1}) and partitioned deployments using the application-specific policy~(Figure~\ref{subfig:tpcc_low_s3_tpcc}), we observe that the \api{lsd}-aware OCC variant incurred in marginal overhead.
In the partitioned deployment using the application-agnostic policy~(Figure~\ref{subfig:tpcc_low_s3_hash}), the overhead becomes more pronounced~($\approx1.25$--$1.5\times$) due to the additional communication round needed to commit some distributed transactions.
However, at high load the \api{lsd}-aware variant managed to achieve similar to better performance.
In contrast, the \api{lsd}-aware 2PL exhibits worse performance than either protocol using the standard interface.

We conclude that the \api{lsd}-aware OCC protocol is not only the best of the \api{lsd} variants, but also the best solution when either using a single database or a partitioned database with a partitioning scheme that allows for committing distributed transactions without incurring in additional communication rounds.
Even with additional communication rounds, \api{lsd} is able to reap better performance under contention, while still providing competitive performance when contention is low.
% subsubsection (end)
% subsection (end)
\subsection{Microbenchmarks} % (fold)
\label{sub:evaluation_micro}
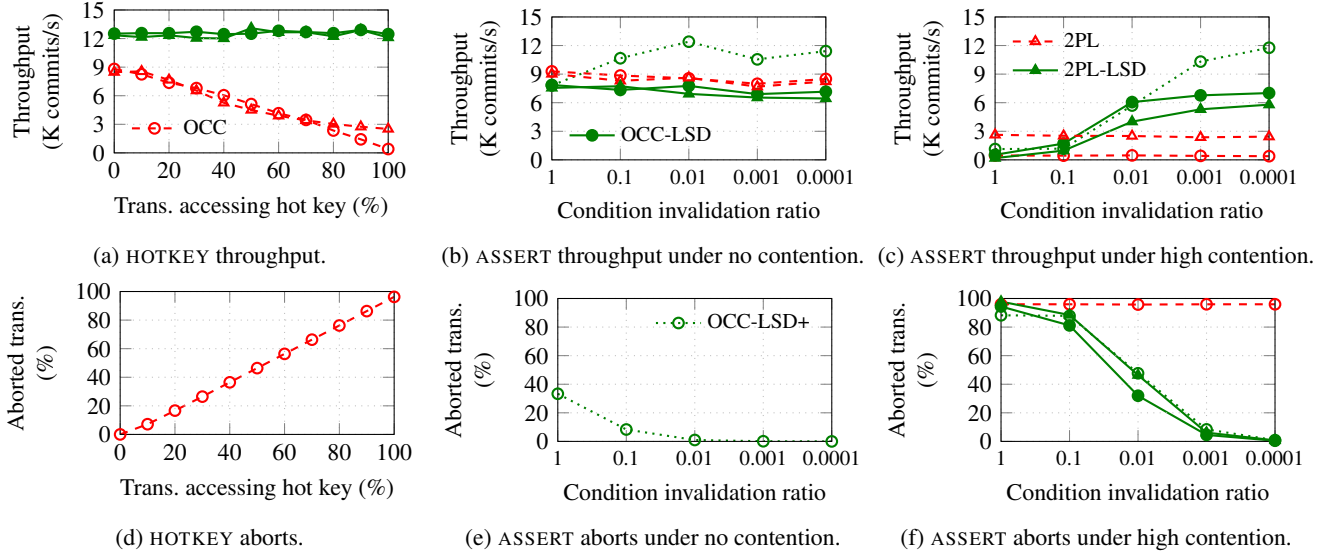
\begin{figure*}[tb]
% throughput
	% hotkey
	\begin{subfigure}[b]{0.33\textwidth}
		\centering
		\begin{tikzpicture}
			\begin{axis}[
				xlabel={Trans. accessing hot key ($\%$)},
				xlabel near ticks,
				x label style={align=center,font=\small},
				x tick label style={font=\small},
				ylabel={Throughput \\ (K commits/s)},
				ylabel near ticks,
				y label style={align=center,font=\small},
				y tick label style={font=\small},
				xmin=0, xmax=100,
				ymin=0, ytick={0, 3000, 6000, 9000, 12000, 15000}, ymax=15000,
				yticklabels={0, 3, 6, 9, 12, 15},
				scaled y ticks=false,
				ymajorgrids=true, yminorgrids=true, 
				xmajorgrids=true, xminorgrids=true,
				grid style={dotted,gray!50},
				axis on top,
				legend cell align={left},
				legend entries={{OCC}},
				legend columns=1,
				legend style={fill=none,draw=none,font=\footnotesize},
				legend pos=south west,
				width=0.9\linewidth,
				height=0.6\linewidth
			]
				\addplot [mark size=2, thick, mark options={solid}, dashed, red!100!black, mark=o] table {
0 8808.33333333333
10 8247.33333333333
20 7364.66666666667
30 6801.66666666667
40 6049.33333333333
50 5125
60 4163.33333333333
70 3461
80 2347.33333333333
90 1421.33333333333
100 402
				};
				\addplot [mark size=2, thick, green!50!black, mark=*] table {
0 12517.3333333333
10 12569
20 12562.6666666667
30 12713.6666666667
40 12462
50 12510.6666666667
60 12815
70 12689
80 12565
90 12905.6666666667
100 12467.3333333333
				};
				\addplot [mark size=2, thick, mark options={solid}, dashed, red!100!black, mark=triangle] table {
0 8479.33333333333
10 8553.33333333333
20 7659
30 6538
40 5258.66666666667
50 4505.66666666667
60 3896.33333333333
70 3412.33333333333
80 3033.66666666667
90 2743.66666666667
100 2513.33333333333
				};
				\addplot [mark size=2, thick, green!50!black, mark=triangle*] table {
0 12342.6666666667
10 12153.6666666667
20 12339.6666666667
30 12048.3333333333
40 12014.3333333333
50 13086
60 12658
70 12666
80 12240
90 12895.3333333333
100 12111
				};
			\end{axis}
		\end{tikzpicture}
		\caption{\api{hotkey} throughput.}
		\label{subfig:hotk_throughput}
	\end{subfigure}
	% assert no contention
	\begin{subfigure}[b]{0.33\textwidth}
		\centering
		\begin{tikzpicture}
			\begin{axis}[
				xlabel={Condition invalidation ratio},
				xlabel near ticks,
				x label style={align=center,font=\small},
				x tick label style={font=\footnotesize},
				ylabel={Throughput \\ (K commits/s)},
				ylabel near ticks,
				y label style={align=center,font=\small},
				y tick label style={font=\small},
				xmin=0, xtick={0, 1, 2, 3, 4}, xmax=4,
				xticklabels={1, 0.1, 0.01, 0.001, 0.0001},
				ymin=0, ytick={0, 3000, 6000, 9000, 12000, 15000}, ymax=15000,
				yticklabels={0, 3, 6, 9, 12, 15},
				scaled y ticks=false,
				ymajorgrids=true, yminorgrids=true, 
				xmajorgrids=true, xminorgrids=true,
				grid style={dotted,gray!50},
				axis on top,
				legend cell align={left},
				legend entries={,{OCC-LSD}},
				legend columns=1,
				legend style={fill=none,draw=none,font=\footnotesize},
				legend pos=south west,
				width=0.9\linewidth,
				height=0.6\linewidth
			]
				\addplot [mark size=2, thick, mark options={solid}, dashed, red!100!black, mark=o] table {
0 9290.33333333333
1 8854.66666666667
2 8555
3 7986.66666666667
4 8493.66666666667
				};
				\addplot [mark size=2, thick, green!50!black, mark=*] table {
0 7858.33333333333
1 7341.33333333333
2 7761.66666666667
3 6901.66666666667
4 7154.33333333333
				};
				\addplot [mark size=2, thick, mark options={solid}, dotted, green!50!black, mark=o] table {
0 7811.33333333333
1 10665.6666666667
2 12402.6666666667
3 10548.6666666667
4 11413.6666666667
				};
				\addplot [mark size=2, thick, mark options={solid}, dashed, red!100!black, mark=triangle] table {
0 9020.66666666667
1 8300.33333333333
2 8621.33333333333
3 7685.66666666667
4 8228.33333333333
				};
				\addplot [mark size=2, thick, green!50!black, mark=triangle*] table {
0 7555.66666666667
1 7727.66666666667
2 6925.66666666667
3 6538
4 6444
				};
			\end{axis}
		\end{tikzpicture}
		\caption{\api{assert} throughput under no contention.}
		\label{subfig:asrt_throughput_hk0}
	\end{subfigure}
	% assert high contention
	\begin{subfigure}[b]{0.33\textwidth}
		\centering
		\begin{tikzpicture}
			\begin{axis}[
				xlabel={Condition invalidation ratio},
				xlabel near ticks,
				x label style={align=center,font=\small},
				x tick label style={font=\footnotesize},
				ylabel={Throughput \\ (K commits/s)},
				ylabel near ticks,
				y label style={align=center,font=\small},
				y tick label style={font=\small},
				xmin=0, xtick={0, 1, 2, 3, 4}, xmax=4,
				xticklabels={1, 0.1, 0.01, 0.001, 0.0001},
				ymin=0, ytick={0, 3000, 6000, 9000, 12000, 15000}, ymax=15000,
				yticklabels={0, 3, 6, 9, 12, 15},
				scaled y ticks=false,
				ymajorgrids=true, yminorgrids=true, 
				xmajorgrids=true, xminorgrids=true,
				grid style={dotted,gray!50},
				axis on top,
				legend cell align={left},
				legend entries={,,,{2PL},{2PL-LSD}},
				legend columns=1,
				legend style={fill=none,draw=none,font=\footnotesize},
				legend pos=north west,
				width=0.9\linewidth,
				height=0.6\linewidth
			]
				\addplot [mark size=2, thick, mark options={solid}, dashed, red!100!black, mark=o] table {
0 460
1 441.333333333333
2 468.666666666667
3 416
4 385.333333333333
				};
				\addplot [mark size=2, thick, green!50!black, mark=*] table {
0 532.666666666667
1 1709.66666666667
2 6062.33333333333
3 6769.33333333333
4 7012.66666666667
				};
				\addplot [mark size=2, thick, mark options={solid}, dotted, green!50!black, mark=o] table {
0 1132.66666666667
1 1175
2 5701.33333333333
3 10315.3333333333
4 11768
				};
				\addplot [mark size=2, thick, mark options={solid}, dashed, red!100!black, mark=triangle] table {
0 2628
1 2526.33333333333
2 2504
3 2368
4 2436
				};
				\addplot [mark size=2, thick, green!50!black, mark=triangle*] table {
0 185.333333333333
1 968.666666666667
2 4012.33333333333
3 5304.33333333333
4 5796
				};
			\end{axis}
		\end{tikzpicture}
		\caption{\api{assert} throughput under high contention.}
		\label{subfig:asrt_throughput_hk100}
	\end{subfigure}
% aborts
	% hotkey
	\begin{subfigure}[b]{0.33\textwidth}
		\centering
		\begin{tikzpicture}
			\begin{axis}[
				xlabel={Trans. accessing hot key ($\%$)},
				xlabel near ticks,
				x label style={align=center,font=\small},
				x tick label style={font=\small},
				ylabel={Aborted trans. \\ ($\%$)},
				ylabel near ticks,
				y label style={align=center,font=\small},
				y tick label style={font=\small},
				xmin=0, xmax=100,
				ymin=0, ymax=100,
				ymajorgrids=true, yminorgrids=true, 
				xmajorgrids=true, xminorgrids=true,
				grid style={dotted,gray!50},
				axis on top,
				legend cell align={left},
				legend entries={},
				legend columns=1,
				legend style={fill=none,draw=none},
				legend pos=north west,
				width=0.9\linewidth,
				height=0.6\linewidth
			]
				\addplot [mark size=2, thick, mark options={solid}, dashed, red!100!black, mark=o] table {
0 0
10 7.15076076688895
20 16.6647628195269
30 26.4440359312652
40 36.4532287320896
50 46.3951681868098
60 56.3607324340269
70 66.3066482340551
80 76.219465607119
90 86.2869168484074
100 96.2554536392862
				};
			\end{axis}
		\end{tikzpicture}
		\caption{\api{hotkey} aborts.}
		\label{subfig:hotk_abort}
	\end{subfigure}
	% assert no contention
	\begin{subfigure}[b]{0.33\textwidth}
		\centering
		\begin{tikzpicture}
			\begin{axis}[
				xlabel={Condition invalidation ratio},
				xlabel near ticks,
				x label style={align=center,font=\small},
				x tick label style={font=\footnotesize},
				ylabel={Aborted trans. \\ ($\%$)},
				ylabel near ticks,
				y label style={align=center,font=\small},
				y tick label style={font=\small},
				xmin=0, xtick={0, 1, 2, 3, 4}, xmax=4,
				xticklabels={1, 0.1, 0.01, 0.001, 0.0001},
				ymin=0, ymax=100,
				ymajorgrids=true, yminorgrids=true, 
				xmajorgrids=true, xminorgrids=true,
				grid style={dotted,gray!50},
				axis on top,
				legend cell align={left},
				legend entries={{OCC-LSD+}},
				legend columns=1,
				legend style={fill=none,draw=none,font=\footnotesize},
				legend pos=north east,
				width=0.9\linewidth,
				height=0.6\linewidth
			]
				\addplot [mark size=2, thick, mark options={solid}, dotted, green!50!black, mark=o] table {
0 33.333053168251
1 8.33274354326151
2 0.9825422021743
3 0.101843334859399
4 0
				};
			\end{axis}
		\end{tikzpicture}
		\caption{\api{assert} aborts under no contention.}
		\label{subfig:asrt_abort_hk0}
	\end{subfigure}
	% assert high contention
	\begin{subfigure}[b]{0.33\textwidth}
		\centering
		\begin{tikzpicture}
			\begin{axis}[
				xlabel={Condition invalidation ratio},
				xlabel near ticks,
				x label style={align=center,font=\small},
				x tick label style={font=\footnotesize},
				ylabel={Aborted trans. \\ (\%)},
				ylabel near ticks,
				y label style={align=center,font=\small},
				y tick label style={font=\small},
				xmin=0, xtick={0, 1, 2, 3, 4}, xmax=4,
				xticklabels={1, 0.1, 0.01, 0.001, 0.0001},
				ymin=0, ymax=100,
				ymajorgrids=true, yminorgrids=true, 
				xmajorgrids=true, xminorgrids=true,
				grid style={dotted,gray!50},
				axis on top,
				legend cell align={left},
				legend entries={},
				legend columns=1,
				legend style={fill=none,draw=none,font=\footnotesize},
				legend pos=north east,
				width=0.9\linewidth,
				height=0.6\linewidth
			]
				\addplot [mark size=2, thick, mark options={solid}, dashed, red!100!black, mark=o] table {
0 95.9436788543606
1 95.954845518148
2 95.7490582049195
3 95.9495508239692
4 95.9524739725326
				};
				\addplot [mark size=2, thick, green!50!black, mark=*] table {
0 94.3855994097826
1 81.1483718458275
2 31.9510893132916
3 4.53394473242083
4 0.463185836593466
				};
				\addplot [mark size=2, thick, mark options={solid}, dotted, green!50!black, mark=o] table {
0 88.2419885964507
1 87.8331784052027
2 47.661636945458
3 8.40190295172982
4 0.882263452051926
				};
				\addplot [mark size=2, thick, green!50!black, mark=triangle*] table {
0 97.801348657349
1 88.4643969855411
2 46.3852527608606
3 6.13322578260296
4 0.72006603794328
				};
			\end{axis}
		\end{tikzpicture}
		\caption{\api{assert} aborts under high contention.}
		\label{subfig:asrt_abort_hk100}
	\end{subfigure}
	\caption{
		Throughput and aborts on the \api{hotkey}~(a,d) and \api{assert} microbenchmarks with no~(b,e) and high~(c,f) contention.
	}
\end{figure*}

In this section we report on microbenchmark results that show the effect of specific workload characteristics on \api{lsd}.

% \noindent{\textbf{Without predicates.}} % (fold)
% \subsubsection{Contention without conditions} % (fold)
% \label{subsub:evaluation_micro_contention_nopredicates}
\paragraph{Contention without conditions.} % (fold
We start by analyzing the effect of contending read-modify-write operations.
To do so, we loaded the database with as many private counters as there were clients, and a single shared counter---the ``hot'' counter.
Transactions consisted of an increment of either the hot counter, according to some probability~$p$, or the respective private counter, with probability $1-p$.
We executed the microbenchmark for various values of~$p$, ranging from~$0\%$~(no contention) to~$100\%$~(all transactions contend).

Figure~\ref{subfig:hotk_throughput} plots the measured throughput as a function of the parametrized contention.
The \api{lsd}-aware protocols are not affected by the parameter because the increments are delayed until commit time, whereas the throughput of the OCC and 2PL protocols decreases when contention increases, as expected, due to aborts in OCC~(Figure~\ref{subfig:hotk_abort}), and transactions blocking when attempting to read the value of the hot counter in 2PL.
At $100\%$ contention, \api{lsd}'s throughput is $\approx5\times$ higher that 2PL and $\approx30\times$ more than OCC.

Even when every transaction only increments its own private counter, the \api{lsd}-aware variants still perform better than their standard counterparts due to the fact that the \api{lsd}'s \api{read} operation does not communicate with the database~(it creates the respective future locally).
\api{lsd} transactions incur in less communication rounds than standard transactions, which translated into an $\approx1.3\times$ increase in throughput.
% subsubsection (end)
% % noindent (end)

% \noindent{\textbf{With predicates.}} % (fold)
%
% \subsubsection{Contention with conditions} % (fold)
% \label{subsub:evaluation_micro_contention_withpredicates}
\paragraph{Contention with conditions.} % (fold
We now analyze the effect of contention in the presence of conditions asserted with the \api{is-true} operation.
Like in the previous microbenchmark, we loaded the database with a set of private counters and a single hot counter.
These counters are initialized with a parametrized value~$n$, and a parametrized percentage of transactions access the hot counter while the remaining access their private counter.
The logic of the transactions consisted of decrementing the  value of the counter if it remained greater than zero, or restoring the its initial value otherwise.
Unlike the previous experiment, in this one we could control the contention that \api{lsd} transactions experienced on the condition: the smaller the  initial value of the counters, the higher the contention, i.e., the condition ``the counter remains greater than zero'' changes at a rate of $\frac{1}{n}$, where~$n$ is the parameterized initial value for the counters.

Figures~\ref{subfig:asrt_throughput_hk0},~\ref{subfig:asrt_throughput_hk100},~\ref{subfig:asrt_abort_hk0}, and~\ref{subfig:asrt_abort_hk100}, depict the throughput and abort percentage of each protocol.
For a scenario with no contention for either \api{lsd} or the standard interface, i.e., each transaction only accesses its private counter, the \api{lsd} variants incur in an overhead of $\approx1.1$--$1.25\times$ when compared to their standard counterparts~(Figure~\ref{subfig:asrt_throughput_hk0}).
This overhead comes from the additional work performed by the \api{is-true} operation, which is not extracting additional parallelism in this experiment because there is no contention.
We also plot a version of the \api{lsd}-aware OCC~(OCC-LSD+) that assumes the counter's value remains greater than zero after the decrement, i.e. it speculates the outcome of the \api{is-true} operation without contacting the database, as discussed in \S\ref{subsub:design_cc_occ}.
The effectiveness of the \api{lsd+} variant depends on the success of its speculation.
As expected, the results in Figure~\ref{subfig:asrt_throughput_hk0} show that the throughput of the \api{lsd+} variant increased when we decreased the condition invalidation ratio, increasing throughput up to $\approx1.3\times$ that of the standard protocols.
The throughput increases because the number of aborts due to failed speculation decreases, as shown in Figure~\ref{subfig:asrt_abort_hk0}.
Only the \api{lsd+} variant aborts in this experiment because each transaction accesses its own private counter.

Next, we examined the situation where all transactions access the hot counter.
This is the worse case scenario for the standard transactions, whereas \api{lsd} transactions can still extract parallelism if the concurrent modifications to the counter do not keep invalidating the condition.
Figure~\ref{subfig:asrt_throughput_hk100} reports the observed throughput as a function of the condition invalidation ratio.
The performance of standard transactions is unaffected by the condition invalidation ratio because standard transactions only deal with concrete values when accessing the counter, so all concurrent transactions conflict: OCC suffers from a high percentage of aborts~(Figure~\ref{subfig:asrt_abort_hk100}) while 2PL suffers from a ``queueing'' effect when acquiring the lock in the \api{read} operation.
Note that in this experiment the results for 2PL are optimal somewhat inflated, because we disabled deadlock prevention for 2PL since transactions only access a single key.
With \api{lsd}, on the other hand, throughput increased as there was more available parallelism to exploit, i.e., updates to the counter that would not make its value fall below~$1$.
In particular, as the abort percentage decreased~(Figure~\ref{subfig:asrt_abort_hk100}), the \api{lsd}-aware variant of OCC (resp.\ 2PL) achieved up to $\approx17\times$~(resp.\ $\approx2\times$) more throughput than its standard counterpart~(Figure~\ref{subfig:asrt_throughput_hk100}).
The \api{lsd+} variant was able to further boost the throughput gains to $\approx30\times$ the performance of OCC.
% subsubsection (end)
% % noindent (end)

% \subsubsection{Number of write functions} % (fold)
% [\textbf{\emph{To do.}}]
% % subsubsection (end)
% \subsubsection{Number of asserted predicates} % (fold)
% [\textbf{\emph{To do.}}]
% % subsubsection (end)

% subsection (end)
% section (end)
\section{Related work} % (fold)
\label{sec:related_work}
\paragraph{Futures and lazy evaluation.}
Sloth~\cite{sloth-sigmod-2014} uses futures and lazy evaluation to reduce the number of network round trips in database-backed applications.
It batches queries at the client until any of the batched query results are needed by the client logic, at which point the batch is sent to the database.
\api{lsd}'s futures achieve the same goal, but \api{lsd} goes further by using futures and lazy evaluation to push application semantics and computation into the database, for the concurrency control protocol to extract more parallelism.
Faleiro et al.~\cite{lazy-sigmod-2014} propose a lazy transaction execution engine.
Transactions must be stored procedures, and the system acknowledges transactions as committed without executing them.
When some transaction needs to observe state that would have been written by some delayed transaction, the delayed transaction and its dependencies are executed.
In contrast, \api{lsd} allows for both stored procedures and the most prevalent client-server execution model (a recent study shows that stored procedures account for less than 10\% of the transactions in most database deployments, and that there are few deployments where all transactions are stored procedures \cite{Pavlo:2017:WDO:3035918.3056096}), and uses futures and lazy evaluation to extract additional read-write parallelism by refining what constitutes a conflict in terms of conditions.
Other proposals~\cite{vboxes-2006,bumper-srds-2013} also delay computation over contended objects until commit but only \emph{if} the values of the objects are not used anywhere else in the transaction's logic.
\api{lsd}'s holistic design of futures and lazy evaluation do not impose this restriction.
%[\textbf{\emph{tv: \cite{promises-pldi-1988}?}}]

\paragraph{Performance under contention.}
Doppel~\cite{doppel-osdi-2014} replicates contended objects across workers to allow parallel commutative updates to each replica, at the expense of preventing the execution of transactions that need to read or perform different update operations.
% Doppel extracts parallelism for certain types of transactions to the detriment of others, while in \api{lsd} any transaction can execute at any time.
\api{lsd} extracts read-write parallelism using futures, conditions, and lazy evaluation instead.
\api{Rococo}~\cite{rococo-osdi-2014} requires programmers to organize transaction logic in pieces that access one or more objects stored on a single partition.
Developers must therefore be aware of the partitioning policy, and the code is tied to a particular policy.
All transactions need to be known in advance to perform complex static analysis.
Callas~\cite{callas-sosp-2015} automates Salt's methodology~\cite{salt-osdi-2014} by requiring transactions to be known in advance to perform static analysis to expose intermediate states to other transactions. 
\api{lsd} improves performance without requiring changes to transaction logic %, is not tied to particular partitioning policies, 
and to know transactions beforehand.

%Callas~\cite{callas-sosp-2015} automates Salt's methodology~\cite{salt-osdi-2014} by requiring transactions to be known in advance to perform static analysis to expose intermediate states to other transactions. The homeostatis protocol~\cite{homeostasis-sigmod-2015} allows distributed databases to execute transactions without coordination across partitions under certain conditions identified in static analysis. \api{lsd} is a simpler solution that extracts concurrency in both centralized and distributed databases but still requires, albeit ``smarter,'' coordination.

Other systems explore the semantics of applications to maintain correctness under non-linearizable executions. Escrow transactions~\cite{escrow-tods-1986} and the demarcation protocol~\cite{demarcationproto-vldbj-1994} maintain global invariants. The homeostatis protocol~\cite{homeostasis-sigmod-2015} allows distributed databases to execute transactions without coordination across partitions under certain conditions identified by static analysis. In contrast, \api{lsd} improves performance under contention while providing linearizability.

%The demarcation protocol~\cite{demarcationproto-vldbj-1994} could be used to efficiently implement the \api{is-true} operation for multi-future conditions using pessmistic techniques.
%[\textbf{\emph{tv: \cite{hightraffic-pods-1982,escrow-tods-1986}?}}]

\paragraph{Performance.}
Silo~\cite{silo-sosp-2013} refine OCC to improve the performance of in-memory databases.
FaRM~\cite{farm-sosp-2015} exploits new hardware functionality in partitioned databases.
\api{lsd} redefines what constitutes a conflict so it is complementary to these proposals.
Sinfonia~\cite{sinfonia-sosp-2007} proposes a restricted form of transactions called minitransactions, whose execution can be piggybacked in the 2PC protocol at the expense of expressiveness, e.g., it is impossible to perform a read-modify-write operation in a single minitransaction.
\api{lsd} does not impose these restrictions on expressiveness, and yet \api{lsd} transactions that do not observe state before attempting to commit are also piggybacked in 2PC.
The \api{is-true} operation resembles warranties~\cite{warranties-nsdi-2014} but \api{lsd}'s design with futures and lazy evaluation extracts concurrency even in cases where warranties are not helpful, such as the example of Figure~\ref{fig:api}.
%\cite{spanner-osdi-2012,calvin-sigmod-2012,f1-vldb-2013,lynx-sosp-2013,orthrus-sigmod-2016}

\paragraph{Concolic execution.}
\api{lsd}'s approach can be interpreted as concolic execution~\cite{cute-esec-fse-2005,klee-osdi-2008} of transactions.
\api{lsd}'s futures are similar to symbolic values, and transactions collect constraints to the possible concrete values futures will resolve to using the \api{is-true} operation.
\api{lsd} develops these concepts in the context of concurrency control to improve performance while still maintaining transactional isolation.
% section (end)
\section{Conclusion} % (fold)
\label{sec:conclusion}
This paper presented \api{lsd}, a refined interface for database transactions.
By allowing transactions to execute their logic over an abstract state and specifying their intent more clearly to the database, the concurrency control protocol can make more informed.
As a consequence, \api{lsd} enables high-per\-for\-mance linearizable transactions even under high contention.
% section (end)

\balance

\bibliographystyle{acm}
\bibliography{references}

\end{document}